\documentclass{emulateapj}

\usepackage{url}
\usepackage{graphicx}
\usepackage{epstopdf}
\usepackage{longtable}
\usepackage{soul,color}

\shorttitle{}
\shortauthors{Sheen et al.}

\begin{document}


\title{Post-merger Signatures of Red-sequence Galaxies in Rich Abell Clusters at $\lowercase{z}\lesssim0.1$}


\author{Yun-Kyeong Sheen\altaffilmark{1, 3},  Sukyoung K. Yi\altaffilmark{1, 4}, Chang H. Ree\altaffilmark{2}, Jaehyun Lee\altaffilmark{1}}


\altaffiltext{1}{Department of Astronomy and Yonsei University Observatory, Yonsei University, Seoul 120-749, Republic of Korea}
\altaffiltext{2}{Korea Astronomy \& Space Science Institute, Daejeon, 305-348, Republic of Korea}
\altaffiltext{3}{Current address: Departamento de Astronom\'{i}a, Universidad de Concepci\'{o}n, 
Casilla 160-C, Concepci\'{o}n, Chile}
\altaffiltext{4}{Corresponding author: yi@yonsei.ac.kr}


\begin{abstract}

We have investigated the post-merger signatures of red-sequence galaxies in rich Abell clusters at $z \lesssim$ 0.1: A119, A2670, A3330 and 
A389. Deep images in $u'$, $g'$, $r'$ and medium-resolution galaxy spectra were taken using MOSAIC 2 CCD and Hydra MOS mounted on a
Blanco 4-m telescope at CTIO. Post-merger features are identified by visual inspection based on asymmetric disturbed features, faint structures, 
discontinuous halo structures, rings and dust lanes. We found that $\sim$ 25\% of bright ($M_r <  -20$) cluster red-sequence galaxies show 
post-merger signatures in four clusters consistently. Most ($\sim$ 71\%) of the featured galaxies were found to be bulge-dominated, and for the 
subsample of bulge-dominated red-sequence galaxies, the post-merger fraction rises to $\sim$ 38\%. 
We also found that roughly 4\% of bulge-dominated red-sequence galaxies interact (on-going merger). 
A total of 42\% (38\% post-merger, 4\% on-going merger) of galaxies show merger-related features. 
Compared to a field galaxy study with a similar limiting magnitude \citep{van05}, our cluster study presents a similar post-merger fraction but a 
markedly lower on-going merger fraction. The merger fraction derived is surprisingly high for the high density of our clusters, where the fast 
internal motions of galaxies are thought to play a negative role in galaxy mergers. The fraction of post-merger and on-going merger galaxies 
can be explained as follows. Most of the post-merger galaxies may have carried over their merger features from their previous halo environment,
whereas interacting galaxies interact in the current cluster in situ. According to our semi-analytic calculation, massive cluster haloes may 
very well have experienced tens of halo mergers over the last 4--5 Gyr; post-merger features last that long, allowing these features to be detected
in our clusters today. The apparent lack of dependence of the merger fraction on the clustocentric distance is naturally explained this way.
In this scenario, the galaxy morphology and properties can be properly interpreted only when the halo evolution characteristics are understood 
first.
\end{abstract}





\section{Introduction}

The formation of massive early-type galaxies in the universe is still in question. 
The hierarchical galaxy formation scenario is widely accepted at present, which predicts that massive galaxies form through hierarchical galaxy
mergers. If each galaxy merger induces star formation and consequent stellar mass growth in the united galaxy, it would lead to a large scatter 
in the age and metallicity of stellar populations in massive early-type galaxies. However, the observational characteristics of early-type galaxies,
such as their red colors tightly correlated in optical color-magnitude relations (CMRs) and their high $\alpha-$elements ratios, imply that most of
the stellar contents in massive early-type galaxies formed in a short timescale at an early epoch ($z >$ 1). Furthermore, the red-sequence in the
optical CMR appears to be established by $z \sim$ 1 \citep{tan05}, and recent observations with 8--10m-class telescopes reveal the appearance
of massive red-sequence galaxies ($M \gtrsim 10^{11} $M$_{\sun}$) at $z \sim$ 2 -- 3 \citep{kod07,dro08,kan09}. These observational results 
suggest that most massive early-type galaxies had almost completed their star formation and mass aggregation by $z \sim$ 1 and then virtually 
passively evolved.

There are other observational clues which indicate a significant increase in the stellar mass density in massive red-sequence galaxies since 
$z \sim$ 1, which is not allowed according to the passive evolution of blue galaxies alone. 
\citet{bel04} showed that the $B$-band luminosity density of massive red-sequence galaxies does not evolve much in the redshift range of 
$0 < z \leq 1.1$. Because the $B$-band light should dim as stars get old, they argued that some young populations should be provided regularly 
from blue galaxies to the red-sequence during this period. However, the simple fading of massive blue galaxies cannot be applied, as the 
brightest red galaxy is always brighter than the brightest blue galaxy throughout the redshift range. After all, they suggested that galaxy mergers 
are an important process in the formation of luminous red galaxies since $z \sim$ 1. The result was supported again by \citet{fab07}.

Residual star formation (RSF) in early-type galaxies provides another clue about galaxy formation. RSF has been extensively studied, recently 
using the UV data from space telescopes such as the Galaxy Evolution Explore ($GALEX$) and the Hubble Space Telescope ($HST$).
The latest work on the UV upturn phenomenon of early-type galaxies showed that RSF fraction among cluster elliptical galaxies is as high as 
$\sim$ 30\%; it is even higher in the field environment \citep{yi11}.
Several studies have claimed that the RSF in early-type galaxies is most likely related to galaxy mergers or interactions \citep{kav07,kav10b}.
Because the UV bright phase of a young stellar population lasts only $\sim$ 1 Gyr, we can estimate that the RSF detected at less than 
$z =$ 0.1 is stimulated at a relatively low redshift, $z \sim$ 0.2 -- 0.3. 
Again, this is indirect evidence of the substantial frequency of galaxy mergers given a low redshift.

\citet{van05} showed tidally disturbed features around field elliptical galaxies at $z \sim$ 0.1 with optically deep images from NDWFS (NOAO
Deep Wide Field Survey) and MUSYC (MUlti-wavelength Survey of Yale and Chile). 
That paper suggested that $\sim$ 70\% of field elliptical galaxies were assembled through dry mergers (i.e., mergers of gas-poor, 
bulge-dominated systems) in the recent epoch. 
\citet{kav10a} also found disturbed features as well as dust lanes from 25\% of luminous early-type galaxies ($M_r < -20.5$, $z <$ 0.05) in the
`Strip 82' fields of Sloan Digital Sky Survey (SDSS) data, which have much longer integration times than general SDSS fields. 
These are very meaningful results because mass growth via continuous galaxy mergers were up to that point mostly supported observationally 
by statistical analyses of photometric colors instead of by visual evidence of galaxy mergers. 
Moreover, the merger fraction is often estimated from galaxy pair fractions, considering them as pre-mergers \citep[e.g.,][]{bun09,der09}. 
Although \citet{van05} and \citet{kav10a} did not deal with complete volume-limited samples, they searched a substantial number of galaxies 
with unprecedentedly deep optical images and presented direct evidence of galaxy mergers related to the formation of red, early-type galaxies.

Our interest moved to galaxy clusters, which are dominated by massive early-type galaxies. It has been expected that galaxy clusters are not a
likely environment for frequent merger events to take place due to the high peculiar motions of the galaxies within the cluster. 
Moreover, it is commonly held that massive galaxies in clusters formed faster at an earlier epoch than did galaxies in a field environment 
(Gunn \& Gott 1972; Dressler 1980; Tanaka et al. 2005 and references in it). Therefore, it was presumed that it would be difficult to find 
post-merger signatures from massive early-type galaxies in galaxy clusters. Naturally, there has been less effort to find post-merger 
signatures among cluster galaxies.

To find post-merger signatures of early-type galaxies in galaxy clusters, we carried out optical deep imaging and multi-object spectroscopic 
observations of \st{four} rich Abell clusters at 0.04 $< z <$ 0.11 with the Blanco 4-m telescope at the Cerro Tololo Inter-American Observatory 
(CTIO). We picked four rich Abell clusters (A119, A2670, A3330 and A389) as our targets. They have been observed by GALEX at least in the 
medium-depth imaging mode (MIS: Medium Imaging Survey) and are optimally visible from the CTIO. Their GALEX NUV exposure times range 
from $\sim$ 1 hours for A119 to $\sim$ 17 hours for A3330. UV light is very sensitive to the existence of young stellar populations and thus plays 
an important role in the study of the recent star formation history that may have been caused by mergers. 
The UV properties will be published in an upcoming paper following this one, in which we initially present the optical colors and morphologies 
of cluster galaxies. We performed a visual inspection of all the red-sequence galaxies brighter than $M_{r'} = -20$ in each galaxy cluster using 
deep optical images. This result will enable us to compare the recent galaxy merger histories between cluster and field environments. 
Morphological indexes such as the bulge-to-total ratio and asymmetry of the galaxies were also measured to examine their morphological 
characteristics.

In this paper, the observations and data reductions are presented in Section~\ref{sec:obs} and Section~\ref{sec:data}. Section~\ref{sec:sample} 
describes the galaxy sample selection process and Section~\ref{sec:vi} shows the scheme and the result of the visual inspection using the deep 
optical images. Morphological examination will be presented in Section~\ref{sec:mi}. We will conclude this paper with a summary and discussion
in Section~\ref{sec:disc}.

\begin{figure}
\center{
\includegraphics[scale=0.4]{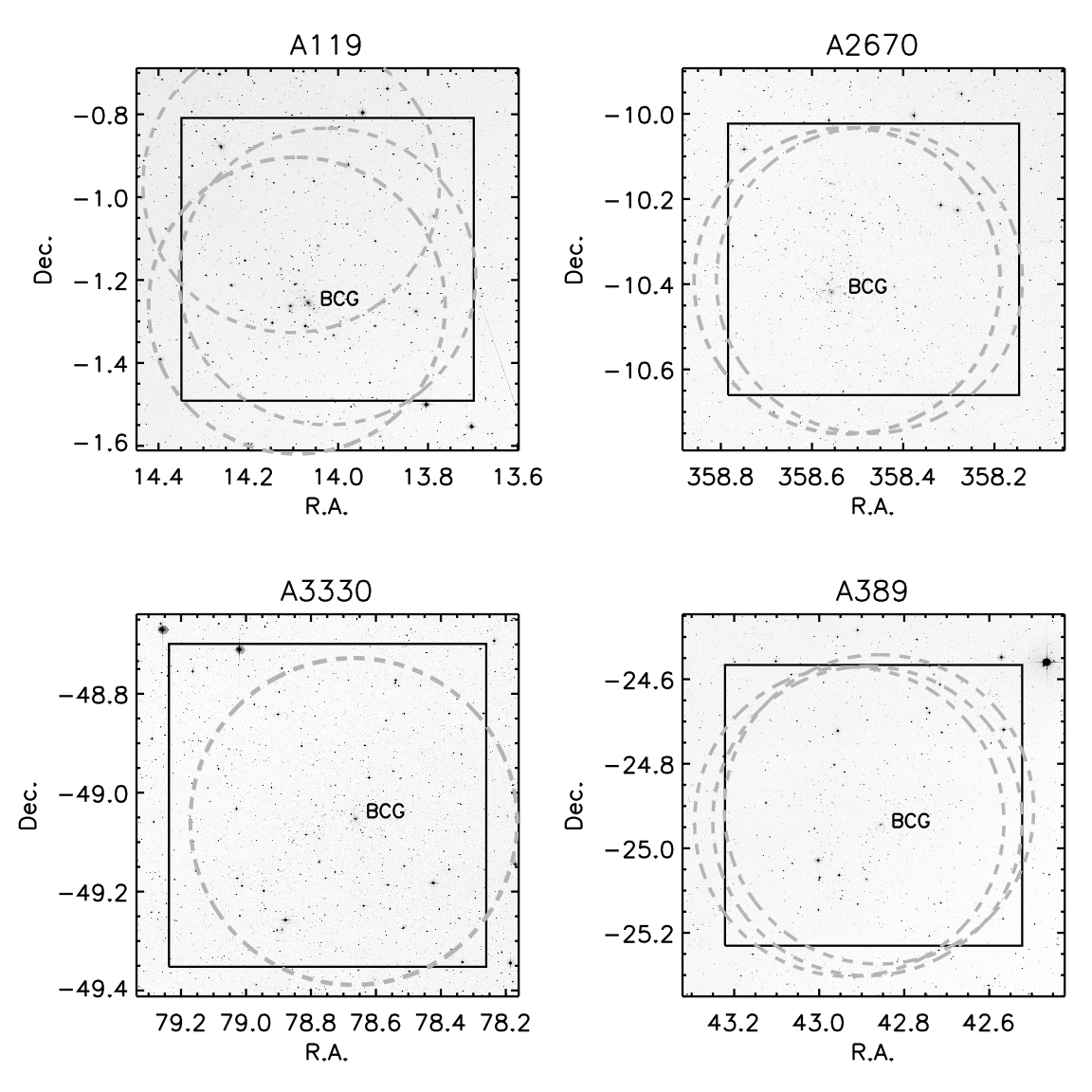}
\caption{The field-of-views of the MOSAIC2 and Hydra data are superimposed over the DSS image for each cluster. The square in each panel 
indicates the field of the stacked deep image while the gray-dashed circles represent the observed Hydra FOVs. Their consistent FOVs enabled 
us to perform an almost complete spectroscopic survey of the cluster galaxies.\label{field}}
}
\end{figure}

\section{Observations}
\label{sec:obs}

\begin{deluxetable*}{cccccccc}
\tabletypesize{\footnotesize}
\tablewidth{0pt}
\tablecaption{CTIO observations summary \label{obssum}}
\tablehead{
\colhead{Cluster ID} & \colhead{RA(J2000)} & \colhead{Dec(J2000)} & \colhead{Instrument} & \colhead{Filter} &
\colhead{$t_{exp}$ (sec)\tablenotemark{a}} & \colhead{\# of pointing\tablenotemark{b}} & \colhead{Date}
}
\startdata
Abell 119 & 00:56:21.4 & -01:15:47 & MOSAIC 2 & $u'$ & 6000 / 1200 & 5 & 27,29 Oct. 2008 \\
($z\sim0.044$) & & & & $g'$ & 5760 / 120 & 8 \\
 & & & & $r'$ & 5760 / 120 & 8 \\
 & & & Hydra & & 9000 & 4 & 6--11 Nov. 2008 \\[6pt]
Abell 2670 & 23:54:10.1 & -10:24:18 & MOSAIC 2 & $u'$ & 4800 / 700 & 6 & 13,15 Nov. 2007 \\
($z\sim0.076$) & & & & $g'$ & 5100 / 200 & 6\\
 & & & & $r'$ & 6000 / 60 & 6\\
 & & & Hydra & & 5400 & 3 & 27--29 Nov. 2007 \\[6pt]
Abell 3330 & 05:14:40.0 & -49:03:15 & MOSAIC 2 & $u'$  & 4600 / 460 & 10 & 26 Nov. 2006 \\
($z\sim0.089$) & & & & $g'$ & 5000 / 200 & 10 &  \\
 & & & & $r'$  & 6150 / 100  & 15 & \\
 & & & Hydra & & 14400 / 3600\tablenotemark{c} & 3/6 & 27--29 Nov. 2007 \\
 & & & & & & & / 1--3 Dec. 2006 \\[6pt]
Abell 389 & 02:51:31.0 & -24:56:05 & MOSAIC 2 & $u'$ & 6000 / 1200 & 5 & 27,29 Oct. 2008 \\
($z\sim0.113$) & & & & $g'$ & 7200 / 1200 & 6 \\
 & & & & $r'$ & 7200 / 600 & 6 \\
 & & & Hydra & & 10800 & 3 & 6--11 Nov. 2008 
\enddata
\tablenotetext{a}{~The first numbers are the total exposure time of the stacked images and the second numbers represent the short exposure 
images that were analyzed.}
\tablenotetext{b}{~The numbers indicate the number of dithering points for imaging and the number of fiber configurations for 
spectroscopy for each target field.}
\tablenotetext{c}{~Spectroscopic redshifts of some A3330 galaxies were complemented by the spectra taken in 2006.} 
\end{deluxetable*}

Deep optical images of the Abell clusters were taken with the MOSAIC 2 CCD mounted on the Blanco 4-m Telescope at CTIO. 
Medium-resolution spectra ($R \sim$ 1000) of galaxies in the cluster fields were acquired with Hydra, a multi-object spectrograph on the Blanco. 
Observations were performed three times through 2006 -- 2008. The target clusters and acquired data are summarized in Table~\ref{obssum}. 
Only the short exposure image ($t_{exp} =$ 600 sec) of A389 was taken separately in early 2009 with the same telescope and instrument. 
MOSAIC 2 consists of eight 2K$\times$4K CCDs (therefore, the overall dimension is 8K$\times$8K pixels), and Hydra utilizes 
a 2K$\times$4K CCD. The pixel scale of MOSAIC 2 is 0.24$\arcsec$/pixel after the astrometry is done. MOSAIC 2 has a $36\arcmin\times36\arcmin$ field-of-view (FOV) 
and Hydra has a circular FOV that is $40\arcmin$ in diameter. The FOVs of MOSAIC 2 and Hydra are plotted over Digitized Sky Survey images 
in Figure~\ref{field} for each cluster. The square in each panel indicates the region of the stacked deep image while the gray-dashed circles 
represent FOVs of Hydra observations. Their consistent FOVs enabled us to perform an almost complete spectroscopic survey of the cluster
galaxies effectively.

All of the deep images were taken on dark nights (moon illumination $<$ 0.4, the average is 0.17) while the spectroscopic observations were 
performed on bright nights. The imaging nights were usually photometric and the average seeing was less than $1\arcsec$ during the entire 
period of observation. We obtained the deep images with $u'$, $g'$, $r'$ filters in the Blanco telescope. In order to fill the gap between the eight 
chips in the MOSAIC 2 CCD, more than five points of dithering was applied for each filter. The maximum number of dithering points was 15 for 
the $r'$ band deep image of Abell 3330. The $u'g'r'i'z'$ standard stars in Chandra Deep Field South\citep[CDF-S,][]{smi03} were also observed 
every night to standardize the photometric data.

Hydra observations were performed with a KPGL3 grating centered at 5205\AA{}. A blocking filter, GG385, was also used to suppress the bluer 
wavelength below 3850\AA. The resolving power was $R \sim 1100$ at 5500\AA{} and the dispersion was 1.40\AA{}/pixel after binning two pixels
along the dispersion axis. Calibration frames were taken with a ``Penray'' lamp for wavelength calibration and with a ``Quartz'' lamp as an 
aperture reference. We used the 138 large fibers available in Hydra. The fiber width was 300 $\micron$, which correspond to $2\arcsec$ in the
sky. In a fiber configuration, $\sim$ 100 fibers were assigned for galaxies while $\sim$ 10 fibers were allocated to blank sky for the sky spectra. 
Three to four fiber configurations were designed for each galaxy cluster for the possible cluster members not to be missed by fiber 
collisions and the limited number of fibers. Every fiber configuration was also taken through three separate exposures for median combine for 
data reduction purposes.

\section{Data}
\label{sec:data}

\subsection{MOSAIC 2 photometric data}
\label{sec:pdata}

\subsubsection{Data reduction}

The MOSAIC 2 data are mainly reduced using the {\tt mscred} package in IRAF. All of the calibration data described in this section were obtained 
for every run and were applied to the object frames observed during the same run. First, we updated the characteristic information of each CCD
chip. New crosstalk files were generated with the {\tt xtcoeff} task using shallow object frames which have saturated stars upon a small sky 
background for the CCD pairs of MOSAIC 2. Doing this corrects the ghost images of saturated stars reflected onto each other between a pair of 
CCDs. Gains and readout noises of eight chips were computed with the {\tt mscfindgain} task. In addition, the bad pixel mask of each chip was 
revised manually via visual inspection. We then updated all of the FITS files of the same observing run with the new information. 
An IRAF task {\tt ccdproc} handled all of the pre-processes at once, such as crosstalk correction, bad pixel mask correction, overscan strip 
correction, image trim, bias correction, and flat-fielding. The combined dome flats were used for the flat-fielding of $g'$, $r'$ band images, while 
the combined sky flat was used for the $u'$ band images.

Once the pre-process was done, we found a smooth background gradient in the $g'$, $r'$ band deep images. Illumination correction was carried 
out for each band using the dark-sky flats generated from the object-subtracted deep images. After illumination correction, the deep images 
showed $\pm$2\% deviation in their background over the field of each frame. Cosmic rays were removed from all of the long-exposure images 
($t_{exp} \geq$ 500 sec) using the {\it lacosmic} routine \citep{van01}. Satellite trails occasionally appeared in the long exposures. 
They were erased using the {\tt satzap} task.

Object frames were stacked to a single deep image in each band. The stacking procedure followed the general guide of the National Optical 
Astronomy Observatory MOSAIC data reduction \citep{jan03}. Astrometry was carried out very carefully for every chip using the {\tt msctpeak}
task in IRAF to achieve positional errors of less than 0.3 $\arcsec$ in RMS (Readers are referred to Valdes 2000 for details).

\subsubsection{Standardization}
\label{sec:std}

Standardization equations were derived for each observing run with $u'g'r'i'z'$ standard stars in CDF-S \citep{smi03}. 
Basically, optical photometry in the $g'$, $r'$ band had to be performed using shallow images ($t_{exp} \sim$ 100 sec), as we had intentionally 
saturated the galactic centers in the deep images to see the faint structures around the galaxies. 
Only for A389, the farthest galaxy cluster among our target clusters, did we use a single deep image ($t_{exp} =$ 1200 sec) and a relatively short 
exposure ($t_{exp} =$ 600 sec) for $g'$ and $r'$ band photometry, respectively. 
The standardization equations were established for the nights in which the short exposures were taken.

The eight chips of MOSAIC 2 CCD have slightly different characteristics in terms of photon efficiency. 
Therefore, great care was required when measuring the brightness from different chips for consistency. 
For this reason, we tried to allocate standard stars to the greatest extent possible in one chip and derived the standardization equations using the 
standard stars in it. Chip 6 of the MOSAIC 2 CCD was dedicated to this work because it has the deepest electron-capacity among the eight chips. 
Stars with small photon-statistical errors ($<$ 0.03 mag) were used in the calculations. As a result, 7 $\sim$ 9 stars were available for the 
standardization of the $g'$, $r'$ band data and 6 stars were available for the $u'$ band data among the 22 standard stars listed in \citet{smi03}. 
Instrumental magnitudes were measured with a fixed aperture diameter, 14$\farcs$86, which is adopted from \citet{smi03}. 
Because the aperture diameter is already larger than 7$\times$FWHM in our data (typical FWHM $\lesssim$ 1$\arcsec$), we did not consider 
aperture correction in the photometry of standard stars.

Standardization equations were established as follows:
\begin{equation}
u' = u'_{inst} + a_{u'} + b_{u'}X_{u'} + c_{u'}(u' - g')
\end{equation}
\begin{equation}
g' = g'_{inst} + a_{g'} + b_{g'}X_{g'} + c_{g'}(g' - r')
\end{equation}
\begin{equation}
r' = r'_{inst} + a_{r'} + b_{r'}X_{r'} + c_{r'}(g' - r')
\end{equation} 
The $u'g'r'$ magnitudes in the above equations are the standard magnitudes and the instrumental magnitudes are defined as $m_{inst} = -2.5$ 
$\times$ log(counts(ADU/sec)) $+$ 25. $X$ is an effective airmass of each object frame. Initially, the basic offsets between the standard 
magnitudes and the instrumental magnitudes were calculated and corrected for the instrumental magnitudes. The magnitudes were then fitted 
versus the airmass. The fitting function provided the relationship with which to determine  the airmass term in the standardization equations. 
In the same manner, we fitted the airmass-corrected magnitudes versus the optical colors and determined the color terms in the equations. 
Weighted-mean values are used throughout the fitting process. The coefficients of standardization equations are presented in Table~\ref{stdeq}.

The standardization equations were applied to the instrumental magnitudes from short exposure images to determine the standard magnitudes 
of galaxies. We then derived the standardization constants for the $u'g'r'$ deep images using the standard magnitudes determined earlier. 
The offsets between the standard magnitudes and the instrumental magnitudes from the deep images were calculated with non-saturated stars 
using aperture photometry. A typical standard deviation of the standardization constants (the average of magnitude offsets) was $\lesssim$ 0.001 mag.

\begin{deluxetable}{ccrrr}
\tablecolumns{5}
\tablewidth{0pt}
\tablecaption{Standardization coefficients \label{stdeq}}
\tablehead{
\colhead{Date} & \colhead{Filter} & \colhead{$a$} & \colhead{$b$} & \colhead{$c$} 
}
\startdata
2006 Nov 26 & $u'$ & -1.478$\pm$0.064 & -0.441$\pm$0.009 & 0.126$\pm$0.048 \\
 & $g'$ & 1.002$\pm$0.012 & -0.168$\pm$0.005 & 0.002$\pm$0.020 \\
 & $r'$ & 0.982$\pm$0.005 & -0.088$\pm$0.002 & -0.018$\pm$0.007 \\[6pt]
2007 Nov 13 & $u'$ & -1.385$\pm$0.086 & -0.399$\pm$0.018 & 0.078$\pm$0.064 \\
 & $g'$ & 1.028$\pm$0.021 & -0.139$\pm$0.012 & -0.049$\pm$0.029 \\
 & $r' $ & 1.001$\pm$0.012 & -0.087$\pm$0.007 & -0.032$\pm$0.014 \\[6pt]
2008 Oct 29 & $u'$ & -1.288$\pm$0.064 & -0.526$\pm$0.011 & 0.095$\pm$0.048 \\
 & $g'$ & 1.119$\pm$0.013 & -0.228$\pm$0.007 & -0.113$\pm$0.020 \\
 & $r' $ & 1.054$\pm$0.006 & -0.155$\pm$0.003 & -0.041$\pm$0.009
\enddata
\end{deluxetable}

\begin{deluxetable}{cccc}
\tablewidth{0pt}
\tablecaption{The average of $k$-correction terms \label{kc}}
\tablehead{
\colhead{Cluster} & \colhead{K$_{u'}$} & \colhead{K$_{g'}$} & \colhead{K$_{r'}$}
}
\startdata
A119 & $-0.20\pm$0.03 & $-0.11\pm$0.02 & $-0.04\pm$0.01 \\ 
A2670 & $-0.29\pm$0.06 & $-0.19\pm$0.03 & $-0.07\pm$0.02 \\
A3330 & $-0.34\pm$0.06 & $-0.24\pm$0.04 & $-0.09\pm$0.02 \\
A389 & $-0.42\pm$0.06 & $-0.32\pm$0.05 & $-0.13\pm$0.03 
\enddata
\end{deluxetable}

\subsubsection{Photometry}

The optical magnitudes of the galaxies were measured with SExtractor from the MOSAIC 2 data and calibrated as described 
in Section~\ref{sec:std}. Essentially, we adopted Auto\_Magnitude (\texttt{MAG\_AUTO}) of SExtractor \citep{ber96} as the representative galaxy 
magnitude.

Galactic foreground extinction was corrected using the reddening maps from \citet{sch98}. The $E(B-V)$ value applied to each cluster is 
0.039, 0.042, 0.028, and 0.016 for A119, A2670, A3330, and A389, respectively. Finally, $k$-correction was performed by means of simple 
two-population (old plus young) modeling on the UV-optical continua from photometry consisting of $GALEX$ FUV, NUV and CTIO 
$u'$, $g'$, $r'$ bands. In the modeling, the old population has a fixed age of 12 Gyr and the young component varies in age 
(0.01 Gyr $\leqslant t_{young} \leqslant$ 10 Gyr) and mass fraction ($10^{-4} \leqslant f_{young} \leqslant$ 1). Section 3 in \citet{yi11} details 
the procedure. The average of the $k$-correction values the for the cluster red-sequence galaxies are presented for each galaxy 
cluster in Table~\ref{kc}. The $k$-corrected magnitudes are used in all of the analyses in this paper.

The distance moduli used for the absolute magnitudes calculations are 36.25, 37.46, 37.88, and 38.32 for A119, A2670, A3330, and A389, 
respectively.

\subsection{Hydra spectroscopic data}
\label{sec:sdata}
Hydra spectroscopic data were reduced using the {\tt hydra} package in IRAF. First, the 2K$\times$2K images (after binning along the dispersion
axis) were pre-processed to flat-fielding as with the imaging data. After the pre-processes, spectral apertures were traced from the 
two-dimensional images and fiber-to-fiber throughput variation was corrected with a combined sky flat taken in each observing run. 
Wavelength calibration and sky subtraction were also carried out. The procedures after pre-processing were performed via the {\tt dohydra} task. 
All of the spectra were taken in three separate exposures for median combine. To combine the spectra, we normalized each spectrum using the
{\tt continuum} task, after which three spectra of a galaxy were combined into one spectrum with the {\tt scombine} task.

Galaxy radial velocities were measured from the combined spectra using the {\tt fxcor} task in IRAF, which adopts a cross-correlation method 
between a rest-frame galaxy spectral template and a galaxy spectrum \citep{ton79}. The early-type galaxy template and the luminous red galaxy 
template from SDSS were utilized for our radial velocity measurement , as we primarily targeted red-sequence galaxies in the Hydra
observations.

\section{Galaxy sample selection}
\label{sec:sample}

\subsection{Cluster memberships}
\label{sec:r200}

\begin{figure}
\center{
\includegraphics[scale=0.4]{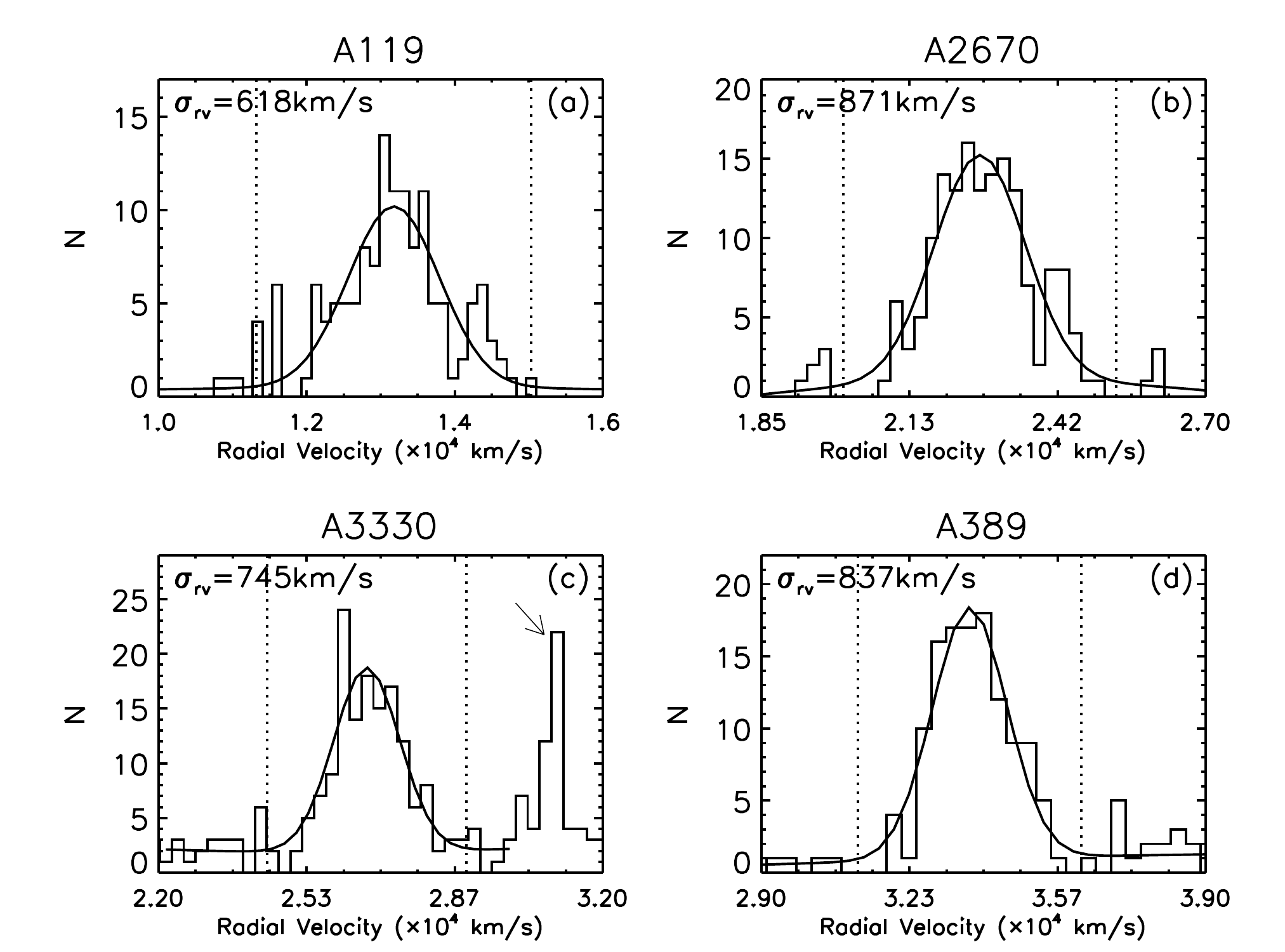}
\caption{Radial velocity histogram of the galaxy clusters. The Gaussian fits are plotted over the histograms. 
The dotted lines indicate the velocities of $v_{r, center}\pm 3 \sigma_{v_{r}}$, with which cluster members are selected.
The arrow in the panel (c) indicates a background cluster of Abell 3330 which appears in Section 4.2.  \label{veldisp}}
}
\end{figure}

\begin{deluxetable*}{ccccccc}
\tabletypesize{\footnotesize}
\tablewidth{0pt}
\tablecaption{Spectroscopic properties of the clusters \label{spectable}}
\tablehead{
\colhead{ID} & \colhead{$v_{r, center}$} & \colhead{$\sigma_{v_{r}}$} & \colhead{$z$} & \colhead{$R_{200}$} & \colhead{$M_{200}$} & \colhead{\# of spectroscopic members\tablenotemark{a}} \\
\colhead{} & \colhead{(km/s)} & \colhead{(km/s)} & \colhead{} & \colhead{(Mpc)} & \colhead{{\scriptsize ($\times10^{14}$M$_{\sun}$)}} & \colhead{}
}
\startdata
A119 &  1.32$\times10^4$ & 618 & 0.044 & 1.53 & 2.6 & 133 (32) \\
A2670 & 2.27$\times10^4$ & 871 & 0.076 & 2.16 & 7.3 & 153 (39) \\
A3330 & 2.67$\times10^4$ & 746 & 0.089 & 1.84 & 4.5 & 130 \\
A389 & 3.37$\times10^4$ & 837 & 0.112 & 2.07 & 6.4 & 114   \enddata
\tablenotetext{a}{The numbers in parenthesis indicate the number of member galaxies identified using SDSS data.}
\end{deluxetable*}

Cluster memberships are determined by spectroscopic redshifts of galaxies. We measured the radial velocity dispersion ($\sigma_{v_{r}}$) 
of the cluster galaxies with the Hydra spectra. Because A2670 and the northern half of A119 are observed by SDSS, we included the SDSS 
spectra of galaxies which we had missed in the Hydra observations. Radial velocity distributions from the combined catalogs (Hydra observations 
+ SDSS archive) are shown in Figure~\ref{veldisp}. The bin sizes are determined as one hundredth of the central velocities of the galaxy clusters. 
The center and dispersion of the radial velocity distributions are determined using a Gaussian fitting IDL code, \texttt{gaussfit}. The fitting result is 
plotted over the histogram with a solid curve. The central velocities and velocity dispersions of the cluster galaxies are presented in 
Table~\ref{spectable}. Cluster memberships are assigned to the galaxies included within $\pm$ 3$\sigma_{v_{r}}$ from the cluster central 
velocity. In Figure~\ref{veldisp}, the dotted lines mark the velocity ranges of the cluster memberships. As a result, we identified 133, 153, 130 
and 114 member galaxies for A119, A2670, A3330 and A389, respectively. Among them, 32 and 39 galaxies are supplemented from the SDSS 
spectroscopic data for A119 and A2670.

To estimate $R_{200}$ from the velocity dispersion, we adopted equation (8) of \citet{car97}. With the assumption of that density profile is a 
singular isothermal sphere in the target clusters, $R_{200}$ can be defined as,
\begin{equation}
R_{200} = \frac{\sqrt{3}}{10}  \frac{\sigma}{H(z)}.
\end{equation}
The $\sigma$ is the characteristic velocity dispersion of a galaxy cluster. In our calculation, we employed the $\sigma_{v_{r}}$ from the Gaussian 
fit of each galaxy cluster for the $\sigma$. The Hubble constant, $H_0$ = 70km s$^{-1}$ Mpc$^{-1}$, is applied to the equation instead of the 
Hubble parameter, $H(z)$, assuming no evolution of $H$ in the redshift range, $0 < z \leq 0.1$. As a result, $R_{200}$ is estimated as 1.53, 2.16,
1.84, and 2.07 Mpc for A119, A2670, A3330, and A389, respectively. The MOSAIC 2 CCD does not cover the entire region of the galaxy clusters. 
The general width of a side of a stacked deep image is $36\arcmin$, which corresponds to 1.9, 3.1, 3.6, and 4.4 Mpc at distances of A119, A2670, 
A3330, and A389, respectively. If it is assumed that a cluster center is located in the center of a deep image, the images cover 62\% (0.95/1.53), 
72\% (1.55/2.16), 98\% (1.8/1.84), and 106\% (2.2/2.07) of $R_{200}$ for each galaxy cluster.

With $R_{200}$, we estimated the mass of the clusters as well. All of the target clusters were more massive than $2\times10^{14}$M$_{\sun}$. 
The mass of A2670 was $\sim 7.3\times10^{14}$M$_{\sun}$, proving that it is one of the most massive galaxy clusters in the universe.

\subsection{Red-sequence galaxies}

\begin{figure}
\center{
\includegraphics[scale=0.35]{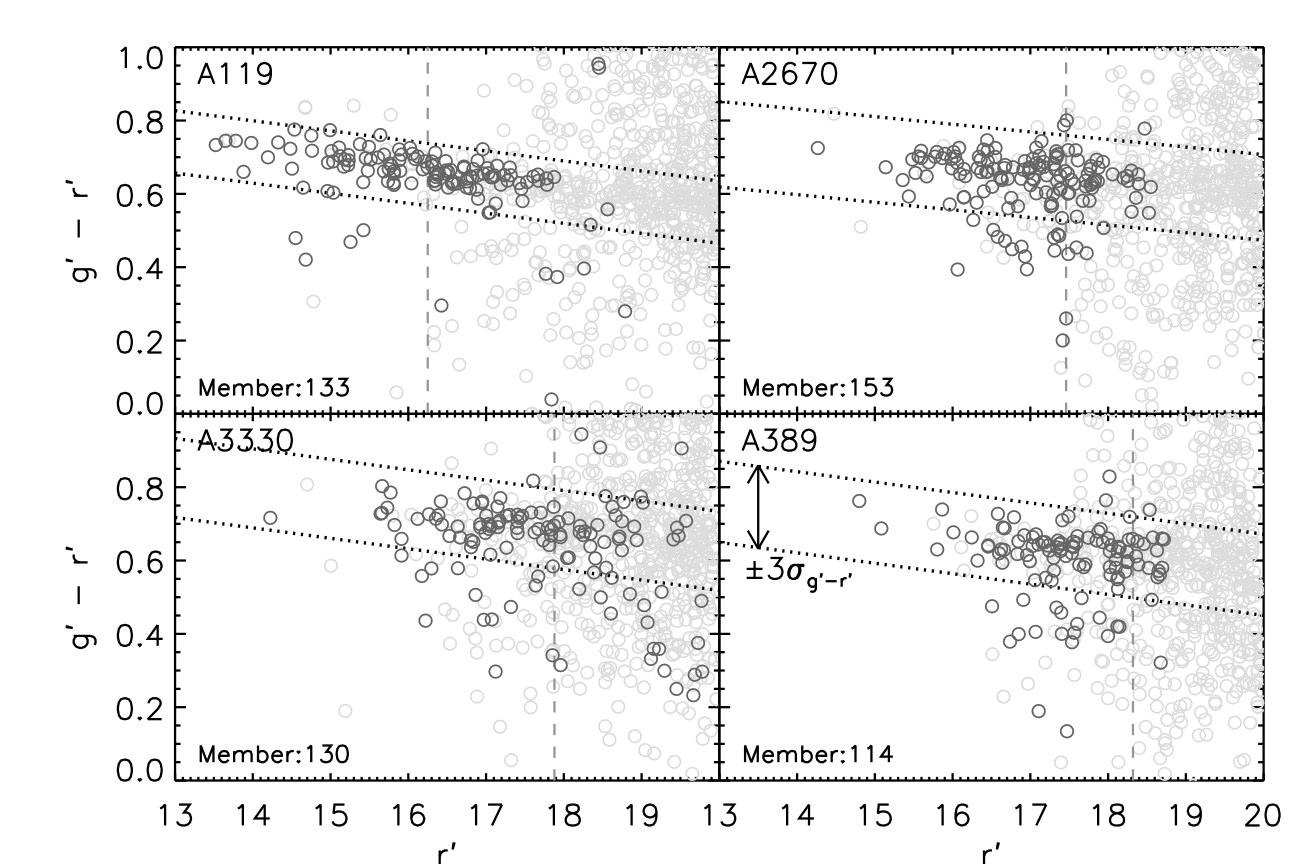}
\caption{Optical color-magnitude relationships of the four target clusters with the MOSAIC 2 data. The dark-gray open circles indicate the
spectroscopic members while the bright-gray open circles present all of the galaxies from the field of deep optical images. 
Red-sequences are defined by iterative linear fitting with the spectroscopic member galaxies. The dotted lines in each panel indicate 
the $\pm 3\sigma$ boundaries from the best fit. We defined the red-sequences with galaxies within the boundaries. The luminosity cut 
($M_{r'} = -20$) for this study is indicated with a gray dashed line in each CMR. \label{cmr_redsq}}
}
\end{figure}

\begin{figure}
\center{
\includegraphics[scale=0.4]{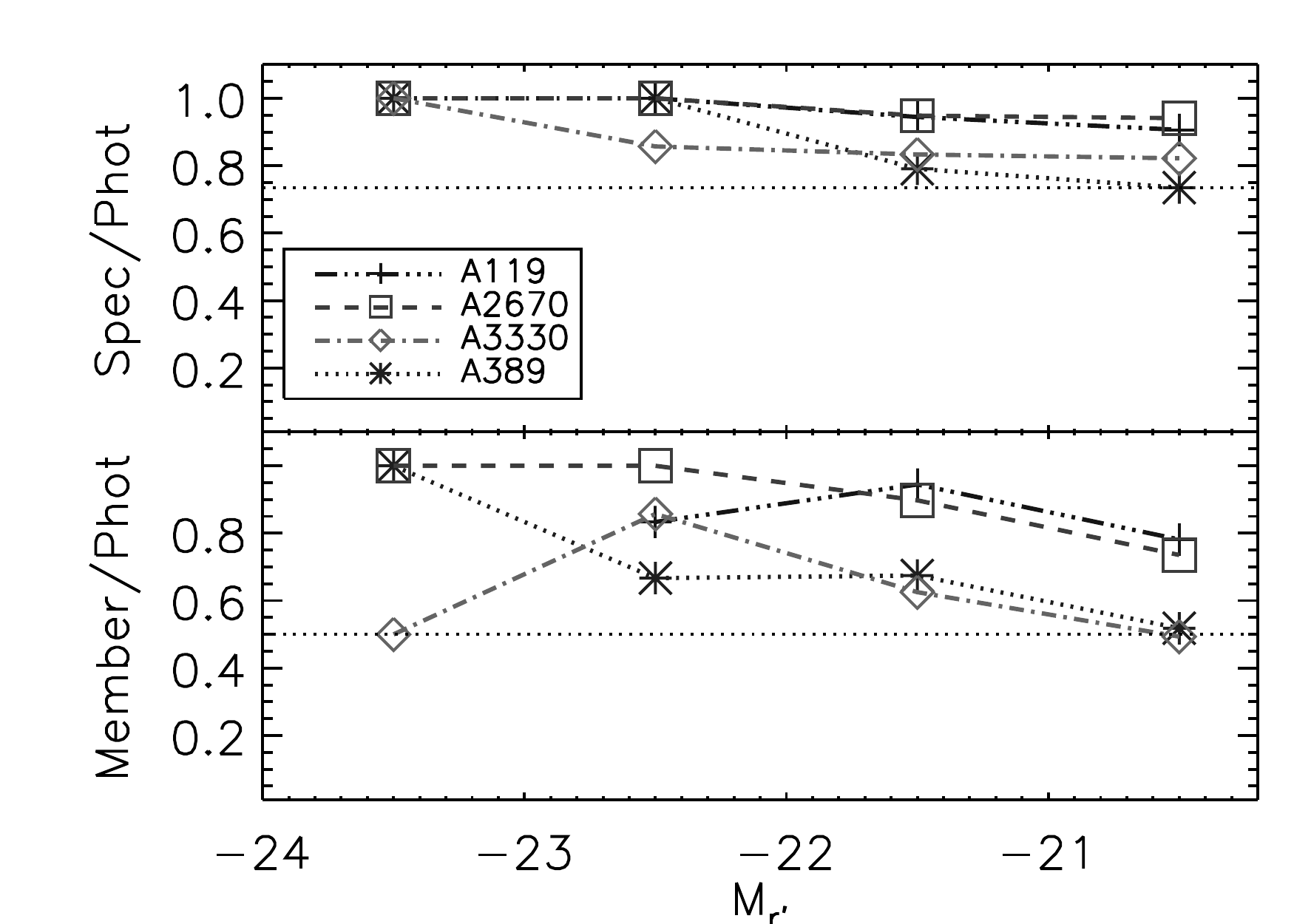}
\caption{The top panel shows the spectroscopic completeness of our data for the red-sequence galaxies along the $r'$ band absolute 
magnitudes. Although the spectroscopic coverage decreases as the galaxies become fainter, still more than 73\% of the $-21 \leq M_{r'} < -20$
photometric red-sequence galaxies are targeted. The bottom panel shows the spectroscopic member fractions among the photometric 
red-sequences along magnitude. The horizontal dotted lines mark 73\% and 50\% in the top and bottom panels, respectively, as the lowest 
fraction in each plot. With this plot, we learn that red-sequences are considerably contaminated by background/foreground galaxies even 
in the cluster fields.  \label{frac_spec}}
}
\end{figure}

Because our interest is focused on the formation of massive red galaxies in galaxy clusters, we defined a red-sequence in each galaxy cluster 
first. The optical color-magnitude relationships of the target clusters are presented in Figure~\ref{cmr_redsq}. In the figure, the dark-gray open 
circles indicate the spectroscopic members while the bright-gray open circles present all of the galaxies from the field of deep optical images. 
Though the red-sequences are distinctively revealed in the CMRs, we empirically defined red-sequences using the spectroscopically identified 
cluster members. However, it was not as simple as visual recognition of the sequences, as a red-sequence has a slight bend near the bright end
of the sequence. Thus, we carried out two-step linear fitting by means of a 2-$\sigma$ clipping method. Initially, a preliminary fit was derived using 
data within a rectangle box which includes galaxies near the bright end of an apparent red-sequence. Once the preliminary fit appears to be close
to the actual sequence, we calculated the standard deviation ($\sigma_{g'-r'}$) of the data to the fit. Linear fitting was then performed again using
the new data set confined by the new $\pm 2 \sigma_{g'-r'}$ color boundaries with $r' < r'_{brightest} + 5$. 
This two-step fitting process allowed us to obtain reasonable fitting results corresponding to what we observed in the CMRs. 
The zero points and slopes for the red-sequences in our targets clusters are $1.107\pm 0.079$ and $-0.026\pm 0.004$, 
respectively. The small scatters are consistent with previous findings of universal red-sequences in galaxy clusters \citep{bow92,sta98}.
Using the final fits, the standard deviations were calculated again with the data within the $\pm 2 \sigma_{g'-r'}$ color boundaries. 
The final values were $\sigma_{g'-r'}$=0.028, 0.039, 0.036, and 0.037 for A119, A2670, A3330, and A389. 
The red-sequences are defined with galaxies within $\pm 3 \sigma_{g'-r'}$ using the final fits and standard deviations. 
The color boundaries are expressed in dotted lines in Figure~\ref{cmr_redsq}.

We attempted to perform a complete spectroscopic survey of the red-sequence galaxies during the Hydra observations. The top panel in 
Figure~\ref{frac_spec} shows the spectroscopic completeness of our data for the red-sequence galaxies along the $r'$ band absolute 
magnitudes. Although the spectroscopic coverage decreases as the magnitude becomes fainter, still more than 73\% of 
the $-21 \leq M_{r'} < -20$ photometric red-sequence galaxies are targeted. The bottom panel shows the spectroscopic member fraction in the 
red-sequences along the magnitudes. With the plot, we learn that red-sequences are considerably contaminated by background/foreground 
galaxies Ð even in the cluster fields. In the case of A2670, though it has very high spectroscopic completeness for the faintest magnitude bin, the 
spectroscopic member fraction is in good agreement with the results from other clusters. Abell 3330 shows the low member fraction for the 
brightest magnitude bin in the bottom panel of Figure~\ref{frac_spec}. It is due to the brightest cluster galaxy of a close background cluster in the 
field. The background cluster is indicated with an arrow in Figure~\ref{veldisp} (c).

The absolute magnitude cut of $M_r < -20$ is applied for the galaxy sample selection because beyond that, spectroscopic survey completeness
is low and visual inspection is difficult for the galaxies at $z \sim$ 0.1. Finally, our galaxy samples are defined by the spectroscopic members in 
the red-sequences; they are also brighter than $M_r = -20$ in each galaxy cluster. We term the galaxy samples `RS$_{sp}$' throughout this paper.

\section{Visual Inspection}
\label{sec:vi}
\subsection{The scheme}
\label{sec:vi_scheme}

Our approach to find post-merger signatures involves a visual inspection of the galaxies with the MOSAIC 2 $r'$ band deep images. 
The faint surface brightness level of $\mu_{r'} \sim$ 30 mag/arcsec$^2$ is acquired in the deep images for all target clusters. The general 
Petrosian diameter along the major axis of the $M_{r'} = -20$ galaxy is $\sim$ 14.4$\arcsec$ ($\sim$ 60 pixels) in the $r'$ band deep images, 
corresponding to $\sim$ 27 kpc at $z =$ 0.1. Generously assuming that a typical seeing is 1 arcsec, the spatial resolution is $\sim$ 1.9 kpc 
at this distance. This can be regarded as the worst condition for our visual inspection.

\begin{figure*}
\center{
\includegraphics[scale=0.8]{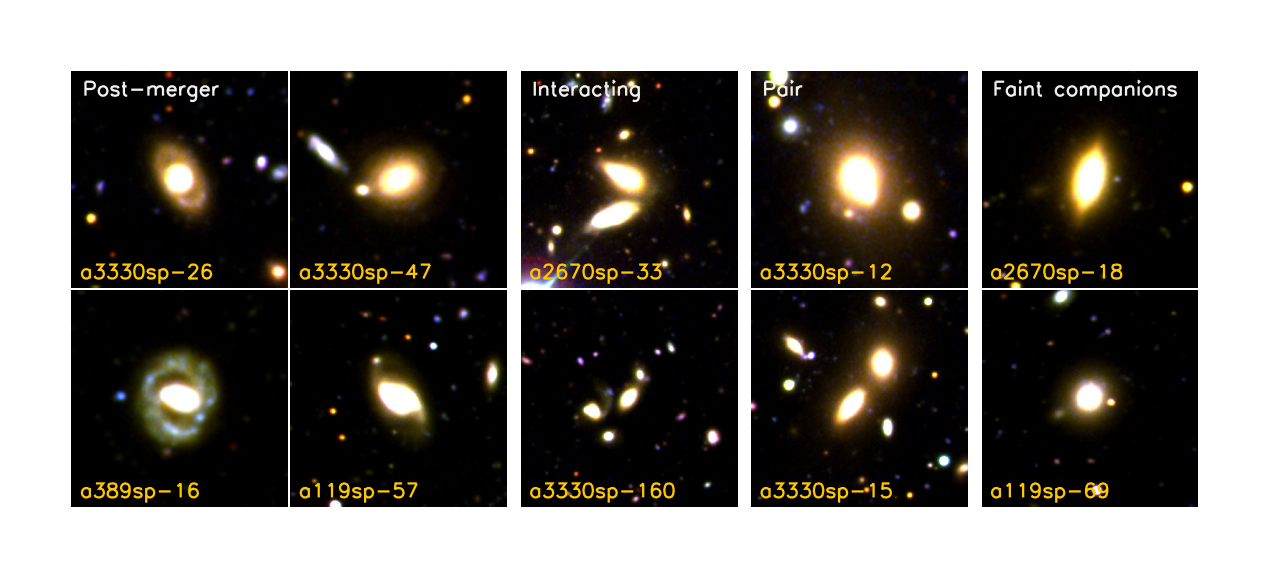}
\caption{This figure shows galaxy examples along the classification scheme. The first two columns show post-merger (PM) galaxy examples and
the examples of interacting galaxies (I), and pair (P), and faint companions (FC) are presented in the following columns.  \label{vi_scheme}}
}
\end{figure*}

We attempted to establish the classification criteria for the visual inspection in a manner faithful to \textit{how the galaxies appear} in the images.
Therefore, the galaxies were classified as ``Elliptical/S0'' (E), ``Post-merger'' (PM), ``Interacting'' (I), ``Pair'' (P), ``accompanying Faint 
Companions'' (FC), and ``Spiral'' (S). ``Elliptical/S0'' denotes the normal elliptical and lenticular galaxies without any disturbed features. 
Additionally, ``Spiral'' indicates normal spiral galaxies. ``Post-merger'' denotes galaxies showing disturbed features, e.g., asymmetric structures, 
faint features, discontinuous halo structure, rings and dust lanes. These are considered as galaxy merger remnants. 
``Interacting'' galaxies also exhibit disturbed features as well as PM galaxies but they are close companions and interact with each other, 
i.e., they are on-going merger candidates. ``Pair'' refers to galaxies with a close companion without any disturbed features. 
We do not know whether they are actual pairs or simply a line-of sight effect of physically separated galaxies. Though we classified these objects, 
we do not include them as interacting systems in this paper. The classification ``accompanying Faint Companions'' was proposed after the 
preliminary visual inspection of the overall populations of the galaxies. In this process, we found many galaxies accompanying faint dwarfish 
objects, mostly showing optical colors similar to those of the main galaxies. The small companions are usually extended along the radial direction
to the center of the main galaxy or along the major axis of the galaxy. We classified them separately, suspecting that they are on-going minor 
merger systems or merger remnants capturing tidal dwarf galaxies that formed during the merger event of the past. Due to the ambiguity of 
identifying the systems, however, we did not include these galaxies in the analyses in this paper, too. Examples of the `PM', `I', `P', `FC' galaxies 
are presented in Figure~\ref{vi_scheme}. It should be noted that only PM and I samples are used in our analysis.

\subsection{The result}

\begin{deluxetable}{ccccc}
\tablecolumns{5}
\tabletypesize{\footnotesize}
\tablewidth{0pt}
\tablecaption{Fractions of the post-merger, interacting, pair, and faint companion galaxies among RS$_{sp}$ \label{vi_result}}
\tablehead{
\colhead{ID} & \colhead{PM} & \colhead{I} & \colhead{P} & \colhead{FC}
}
\startdata
A119 & 27\% (13/48) & 0\% (0/48) &  10\% (5/48) & 10\% (5/48) \\
A2670 & 28\% (25/89) & 8\% (7/89) & 3\% (3/89) & 6\% (5/89) \\
A3330 & 22\% (13/58) & 5\% (3/58) & 9\% (5/58) & 3\% (2/58) \\
A389 & 22\% (17/78) & 4\% (3/78) & 8\% (6/78) & 5\% (4/78) \\
\cutinhead{among the RS$_{sp}$ galaxies within $0.5 R_{200}$}
A119 & 26\% (8/31) & 0\% (0/31) & 10\% (3/31) & 16\% (5/31) \\
A2670 & 29\% (18/63) & 10\% (6/63) & 5\% (3/63) & 8\% (5/63) \\ 
A3330 & 13\% (4/30) & 3\% (1/30) & 7\%(2/30) & 0\% (0/30) \\
A389 & 23\% (10/43) & 5\% (2/43) & 12\% (5/43) & 7\% (3/43) 
\enddata
\end{deluxetable}

\begin{figure}
\center{
\includegraphics[scale=0.45]{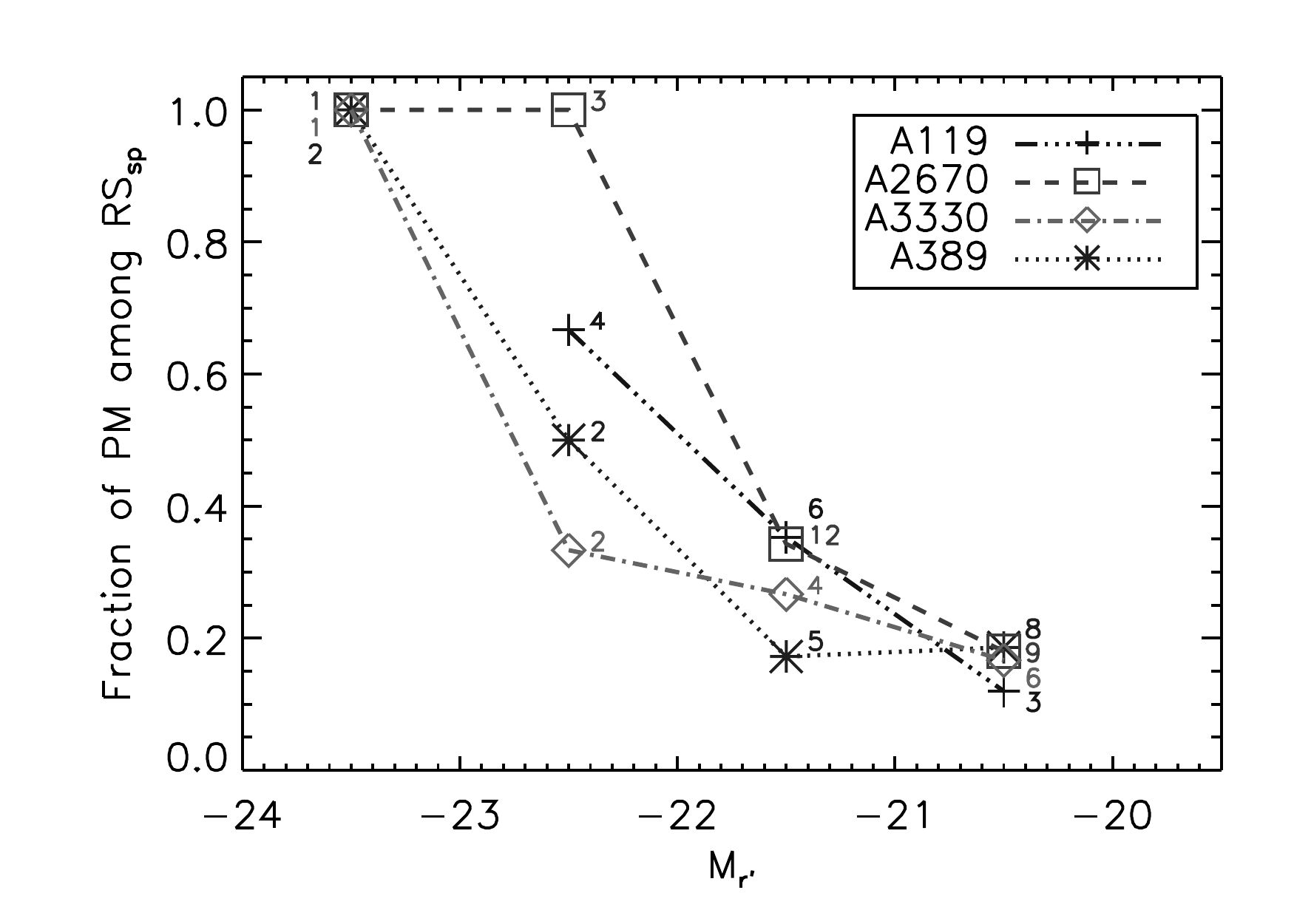}
\caption{The fractions of post-merger galaxies are plotted along the absolute magnitudes. The numbers beside the points indicate the count of the PM galaxies used to derive the fractions. 
Although there is a large spread according to the small values for the bright galaxies, the overall tendency is consistent between all four target 
clusters. This figure shows that post-merger features are more common as red-sequence galaxies become brighter in a cluster environment. 
\label{vi_hist}}
}
\end{figure}

\begin{figure*}
\center{
\includegraphics[scale=0.8]{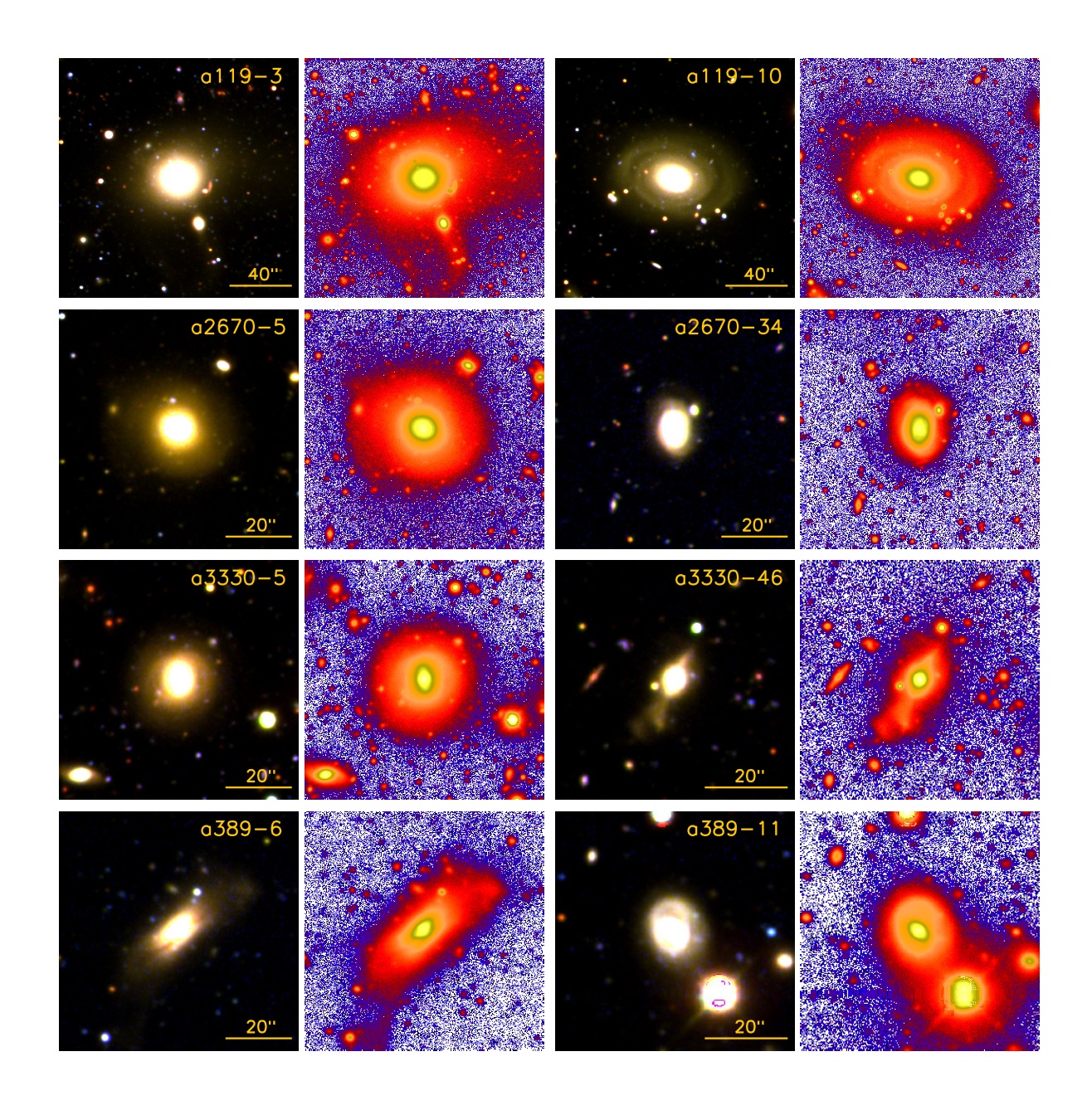}
\caption{Post-merger galaxy samples from four Abell clusters. The first column of each galaxy is a pseudo-color image generated with $u'g'r'$ 
deep images. The second column is a surface brightness map of the galaxy, revealing faint features which do not appear in the color image. 
\label{example}} 
}
\end{figure*}

\begin{deluxetable*}{lcrcccccc}
\tablecolumns{10}
\tablewidth{0pt}
\tabletypesize{\footnotesize}
\tablecaption{A catalog of post-merger galaxies. \label{tableall}}
\tablehead{
\colhead{ID} & \colhead{RA} & \colhead{Dec} & \colhead{$r'$} & \colhead{$g' - r'$} & \colhead{B/T} & \colhead{$\mathcal{A}$}
& \colhead{R (Mpc)\tablenotemark{a}} & \colhead{$z$}
}
\startdata
   a119sp-0 &  00:57:02.1 &  -00:52:31 & 13.52 &  0.73 & 0.84 & 0.539 & 1.324 & 0.044  \\
   a119sp-1 &  00:56:16.1 &  -01:15:19 & 13.65 &  0.74 & 0.98 & 0.399 & 0.000 & 0.044 \\
   a119sp-3 &  00:55:18.8 &  -01:16:38 & 13.78 &  0.74 & \nodata & 0.185 & 0.746 & 0.044 \\
   a119sp-5 &  00:56:22.8 &  -01:12:35 & 13.88 &  0.66 & 0.54 & 0.425 & 0.167 & 0.049  \\
   a119sp-9 &  00:56:02.7 &  -01:20:04 & 14.33 &  0.74 & 0.68 & 0.178 & 0.301 & 0.042   \\
  a119sp-10 &  00:55:08.9 &  -01:02:47 & 14.48 &  0.72 & 0.73 & 0.149 & 1.087 & 0.045  \\
  a119sp-22 &  00:55:25.3 &  -01:18:32 & 14.96 &  0.61 & 0.67 & 0.154 & 0.680 & 0.041 \\
  a119sp-31 &  00:57:07.0 &  -01:23:50 & 15.11 &  0.73 & 0.71 & 1.143 & 0.794 & 0.048  \\
  a119sp-34 &  00:55:05.1 &  -01:15:09 & 15.23 &  0.68 & 0.47 & 1.035 & 0.921 & 0.043  \\
  a119sp-35 &  00:56:38.6 &  -01:23:26 & 15.23 &  0.69 & 0.56 & 0.406 & 0.512 & 0.043   \\
  a119sp-49 &  00:55:55.1 &  -01:03:51 & 15.61 &  0.70 & 0.41 & 0.994 & 0.655 & 0.044   \\
  a119sp-51 &  00:55:43.2 &  -01:02:01 & 15.69 &  0.66 & 0.29 & 0.593 & 0.811 & 0.042 \\
  a119sp-57 &  00:55:23.0 &  -01:12:40 & 15.76 &  0.68 & 0.26 & 0.486 & 0.703 & 0.042  \\
  a2670sp-0 &  23:54:13.7 & -10:25:09 & 14.27 &  0.73 & 0.93 & 0.210 & 0.000 & 0.077  \\
  a2670sp-3 &  23:53:40.5 & -10:24:20 & 15.14 &  0.67 & 0.92 & 0.271 & 0.718 & 0.073  \\
  a2670sp-5 &  23:53:47.6 & -10:37:34 & 15.36 &  0.64 & 0.74 & 0.209 & 1.208 & 0.071  \\
  a2670sp-6 &  23:53:46.7 & -10:16:15 & 15.44 &  0.59 & 0.62 & 0.523 & 0.961 & 0.079   \\
 a2670sp-15 &  23:54:05.7 & -10:18:30 & 15.69 &  0.69 & 1.00 & 0.277 & 0.596 & 0.073  \\
 a2670sp-16 &  23:54:21.5 & -10:25:11 & 15.77 &  0.70 & 0.59 & 0.430 & 0.168 & 0.076  \\
 a2670sp-21 &  23:53:32.7 & -10:34:25 & 15.90 &  0.71 & 0.73 & 0.362 & 1.188 & 0.081  \\
 a2670sp-23 &  23:54:10.4 & -10:29:51 & 15.99 &  0.67 & 0.77 & 0.316 & 0.412 & 0.075  \\
 a2670sp-25 &  23:54:14.8 & -10:24:49 & 16.03 &  0.70 & \nodata & 0.483 & 0.037 & 0.070  \\
 a2670sp-26 &  23:54:39.6 & -10:25:23 & 16.01 &  0.71 & 0.68 & 0.628 & 0.558 & 0.077  \\
 a2670sp-34 &  23:53:56.0 & -10:16:13 & 16.14 &  0.59 & 0.11 & 0.660 & 0.857 & 0.074  \\
 a2670sp-36 &  23:53:29.5 & -10:31:43 & 16.16 &  0.65 & 0.38 & 1.126 & 1.106 & 0.073  \\
 a2670sp-38 &  23:54:01.4 & -10:12:48 & 16.31 &  0.58 & 0.25 & 1.336 & 1.095 & 0.078  \\
 a2670sp-44 &  23:54:02.1 & -10:25:59 & 16.40 &  0.61 & 0.75 & 0.556 & 0.259 & 0.071  \\
 a2670sp-51 &  23:53:24.0 & -10:08:51 & 16.45 &  0.66 & 1.00 & 0.914 & 1.764 & 0.075  \\
 a2670sp-53 &  23:54:17.1 & -10:24:56 & 16.45 &  0.72 & 0.75 & 0.270 & 0.077 & 0.081  \\
 a2670sp-75 &  23:54:30.8 & -10:13:24 & 16.72 &  0.56 & 0.84 & 0.390 & 1.075 & 0.076  \\
 a2670sp-78 &  23:54:14.9 & -10:14:09 & 16.78 &  0.67 & 0.44 & 0.319 & 0.946 & 0.075  \\
 a2670sp-79 &  23:53:20.0 & -10:32:16 & 16.84 &  0.57 & 0.25 & 0.614 & 1.308 & 0.074  \\
 a2670sp-88 &  23:54:10.1 & -10:24:25 & 16.93 &  0.69 & 0.98 & 0.459 & 0.099 & 0.080  \\
a2670sp-103 &  23:53:48.4 & -10:24:54 & 17.12 &  0.67 & 0.58 & 0.580 & 0.545 & 0.080  \\
a2670sp-114 &  23:54:51.9 & -10:28:08 & 17.26 &  0.57 & 0.69 & 0.253 & 0.861 & 0.080  \\
a2670sp-141 &  23:54:19.6 & -10:25:54 & 17.38 &  0.61 & 0.75 & 0.214 & 0.143 & 0.074  \\
a2670sp-142 &  23:54:13.3 & -10:07:13 & 17.39 &  0.61 & 0.07 & 1.332 & 1.543 & 0.076  \\
a2670sp-146 &  23:53:26.8 & -10:24:42 & 17.41 &  0.53 & 0.00 & 1.181 & 1.009 & 0.072  \\
  a3330sp-0 &   05:14:39.5 & -49:03:29 & 14.23 &  0.72 & 0.91 & 0.549 & 0.000 & 0.091  \\
  a3330sp-5 &   05:16:06.0 & -48:53:58 & 15.64 &  0.73 & 0.59 & 0.747 & 2.354 & 0.092  \\
  a3330sp-7 &   05:14:29.4 & -49:05:53 & 15.73 &  0.74 & 0.60 & 0.289 & 0.347 & 0.090  \\
 a3330sp-26 &   05:15:29.5 & -49:03:45 & 16.43 &  0.70 & 0.71 & 0.969 & 1.246 & 0.086  \\
 a3330sp-46 &   05:15:44.3 & -48:53:33 & 16.81 &  0.64 & 0.77 & 0.679 & 1.894 & 0.091 \\
 a3330sp-47 &   05:16:03.3 & -49:14:23 & 16.81 &  0.66 & 0.67 & 1.178 & 2.353 & 0.092  \\
 a3330sp-52 &   05:14:46.4 & -49:04:41 & 16.84 &  0.76 & 0.13 & 0.322 & 0.209 & 0.087  \\
 a3330sp-55 &   05:16:10.5 & -48:53:23 & 16.92 &  0.70 & 0.26 & 0.369 & 2.482 & 0.087  \\
 a3330sp-57 &   05:14:38.3 & -49:01:44 & 16.95 &  0.76 & 0.50 & 0.195 & 0.177 & 0.090  \\
 a3330sp-68 &   05:14:47.2 & -48:45:31 & 17.03 &  0.70 & 0.48 & 0.295 & 1.802 & 0.089  \\
 a3330sp-69 &   05:15:17.2 & -49:03:51 & 17.06 &  0.62 & 0.24 & 0.951 & 0.941 & 0.088  \\
 a3330sp-99 &   05:16:09.4 & -48:54:30 & 17.41 &  0.67 & 0.44 & 1.240 & 2.412 & 0.088  \\
a3330sp-100 &   05:14:22.3 & -49:14:54 & 17.40 &  0.73 & 1.00 & 1.006 & 1.215 & 0.087  \\
   a389sp-0 &   02:51:32.7 & -25:04:23 & 14.80 &  0.76 & 0.70 & 0.286 & 0.978 & 0.113  \\
   a389sp-1 &   02:51:24.8 & -24:56:39 & 15.08 &  0.69 & 0.76 & 0.283 & 0.000 & 0.112  \\
   a389sp-6 &   02:51:46.4 & -24:51:09 & 15.87 &  0.74 & 0.87 & 1.207 & 0.945 & 0.110 \\
  a389sp-11 &   02:51:34.7 & -25:02:44 & 16.25 &  0.66 & 0.77 & 1.670 & 0.806 & 0.116 \\
  a389sp-16 &   02:50:44.9 & -24:45:42 & 16.46 &  0.64 & 0.78 & 1.199 & 1.814 & 0.115  \\
  a389sp-19 &   02:50:54.1 & -24:45:24 & 16.58 &  0.73 & 0.13 & 0.893 & 1.669 & 0.117  \\
  a389sp-21 &   02:51:59.2 & -24:52:31 & 16.60 &  0.62 & 0.00 & 1.104 & 1.170 & 0.109  \\
  a389sp-23 &   02:51:13.2 & -24:56:50 & 16.64 &  0.59 & \nodata & 0.596 & 0.355 & 0.110  \\
  a389sp-61 &   02:50:43.7 & -25:00:11  & 17.27 &  0.55 & 0.07 & 0.941 & 1.332 & 0.116  \\
  a389sp-70 &   02:50:59.6 & -24:54:27 & 17.35 &  0.64 & 0.51 & 0.651 & 0.818 & 0.112  \\
  a389sp-73 &   02:51:22.1 & -24:53:27 & 17.39 &  0.63 & 0.12 & 0.826 & 0.401 & 0.109  \\
  a389sp-82 &   02:51:20.0 & -24:53:14 & 17.40 &  0.65 & 0.64 & 1.024 & 0.443 & 0.117  \\
 a389sp-112 &  02:51:40.8 & -24:49:30 & 17.71 &  0.59 & 0.61 & 0.466 & 1.004 & 0.114  \\
 a389sp-163 &   02:50:26.9 & -24:49:20 & 18.04 &  0.64 & 0.87 & 0.970 & 1.988 & 0.109  \\
 a389sp-175 &   02:51:22.5 & -25:07:41 & 18.13 &  0.52 & 1.00 & 1.061 & 1.354 & 0.116  \\
 a389sp-207 &   02:50:44.3 & -24:55:22 & 18.26 &  0.66 & 0.05 & 1.029 & 1.249 & 0.108  \\
 a389sp-217 &   02:51:27.6 & -24:56:40 & 18.28 &  0.72 & 0.19 & 0.429 & 0.086 & 0.108 
\enddata        
\tablenotetext{a}{A projected distance of a galaxy from the BCG in each galaxy cluster.}                                                                                                          
\end{deluxetable*}

\begin{figure}
\center{
\includegraphics[scale=0.45]{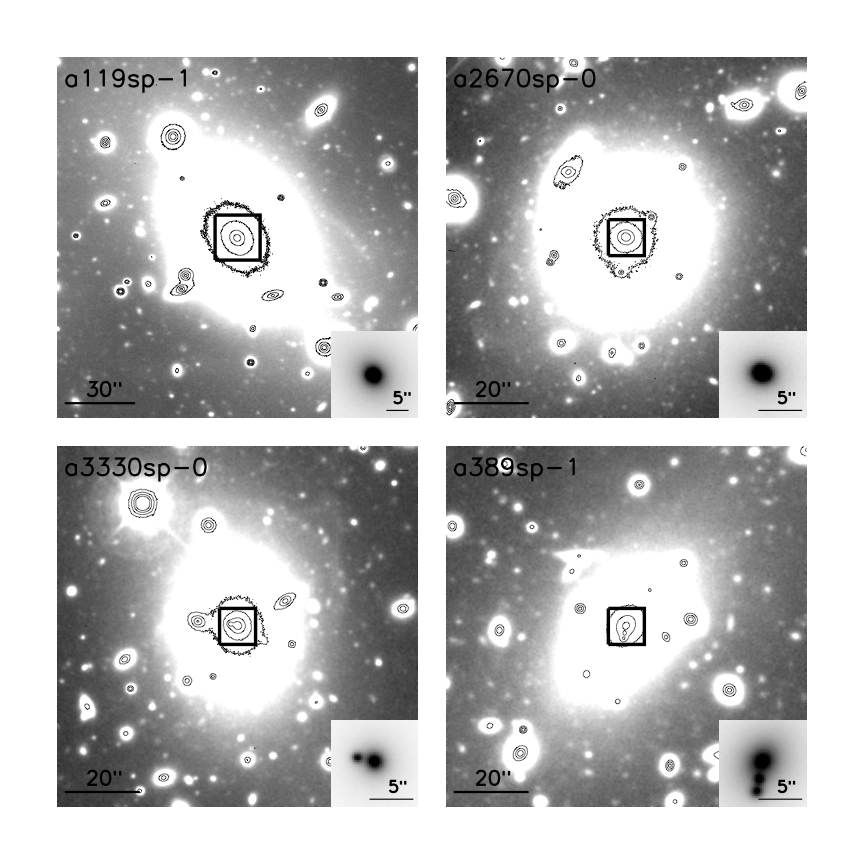}
\caption{The BCGs in our cluster samples are presented with $r'$ band deep images. The contours are drawn for $\mu_{r'} =$ 19, 20, 21, 22 
$mag/arcsec^2$. The inset image in right bottom corner of each galaxy shows the central region indicated by a rectangle box in each image. 
A119, A2670 and A3330 showed faint structures in their halos. Moreover, A3330 and A389 revealed multiple cores in their centers.
This implies that even the BCGs continued violent galaxy mergers until the recent epoch.  \label{example_bcg}}
}
\end{figure}

Visual inspection was performed for the four Abell clusters following the scheme introduced in the previous subsection. In Table~\ref{vi_result}, 
we present the fractions of the post-merger (PM), interacting (I), pair (P) and faint companion (FC) galaxies among the RS$_{sp}$ of each galaxy 
cluster. We found that the $\sim$ 25 $\pm$ 3\% of the RS$_{sp}$ galaxies exhibit post-merger signatures in the galaxy clusters at $z \lesssim$ 
0.1. The fractions of the other classes were usually less than 10\% for each class. If we consider post-mergers and interacting galaxies together, 
the fractions of the disturbed galaxies will be 27\%, 36\%, 28\%, and 26\% respectively in A119, A2670, A3330 and A389. On average, the fraction
of galaxies related to recent galaxy mergers is $\sim$ 30 $\pm$ 4\% in the clusters at $z \lesssim$ 0.1.

The FOVs of the MOSAIC 2 deep images do not cover the entire area within $R_{200}$ for A119 and A2670, as demonstrated in 
Section~\ref{sec:r200}. Therefore, we derived the fractions again with the RS$_{sp}$ galaxies located within $0.5R_{200}$ of each galaxy cluster. 
The results are presented in the bottom half of Table~\ref{vi_result}. Although we applied more consistent areas of clusters, the result did not differ
much from the previous one. The fraction of post-merger galaxies within $0.5R_{200}$ was $\sim$ 24 $\pm$ 4\% on average. 
The value slightly lower than that ($\sim$ 25 $\pm$ 3\%) in the previous sample is most likely a result of A3330 showing a distinctively low 
post-merger fraction within the half of $R_{200}$.

While most of the fractions in a certain class show similar values between the clusters, it is interesting that no interacting galaxies were found 
among the bright red-sequence galaxies of A119. This may have been affected by the relatively small sky coverage of A119 ($\sim 0.6R_{200}$) 
by the MOSAIC 2 deep images. Because galaxy interactions would be more plausible at cluster outskirts, it is possible that we missed some 
galaxies out of the FOV. Another possible reason is that A119 is the smallest galaxy cluster among our target clusters and therefore includes 
fewer bright galaxies. The derived size and mass of the clusters are presented in Table~\ref{spectable}. If we assume that the fraction of 
interacting galaxies within $0.5R_{200}$ of A119 may be similar to the value for A3330 ($\sim$ 3\%), which is the second smallest galaxy cluster
among the clusters, we expect only $\lesssim$ 1 galaxies within the area. Therefore, there is a very low chance to find the bright interacting 
galaxies in the given FOV of the A119 data.

The fractions of post-merger (PM) galaxies are plotted along the absolute magnitudes in Figure~\ref{vi_hist}. The numbers beside the data points 
indicate the counts of PM galaxies used to derive the fractions. Although there is a large spread according to the small values for the bright 
galaxies, the overall tendency is consistent between all four target clusters. This figure shows that post-merger features are more common in 
bright red-sequence galaxies in a cluster environment.

Samples of the selected post-merger galaxies are presented in Figure~\ref{example}. In this figure, the first column of each galaxy image set 
is a pseudo-color image generated with $u'g'r'$ deep images. The second column is a $r'$ band surface brightness map of the galaxy revealing 
faint features which do not appear in the pseudo-color images due to the luminous galactic centers.

In fact, the BCGs in all four target clusters revealed extended faint features or multiple nuclei in our deep images. We present four BCGs 
separately in Figure~\ref{example_bcg} with their $r'$ band deep image and the surface brightness contours. The inset images show the center 
of the BCGs indicated by the rectangle boxes. A119, A2670 and A3330 BCGs showed faint structures in their halos. In addition to that, A3330 
and A389 BCGs revealed multiple cores in their centers. This implies that even the BCGs continued violent galaxy mergers until the recent epoch. 

We provide a catalog of the post-merger galaxies in Table~\ref{tableall} with
their observational properties.
 
\begin{figure}
\center{
\includegraphics[scale=0.45]{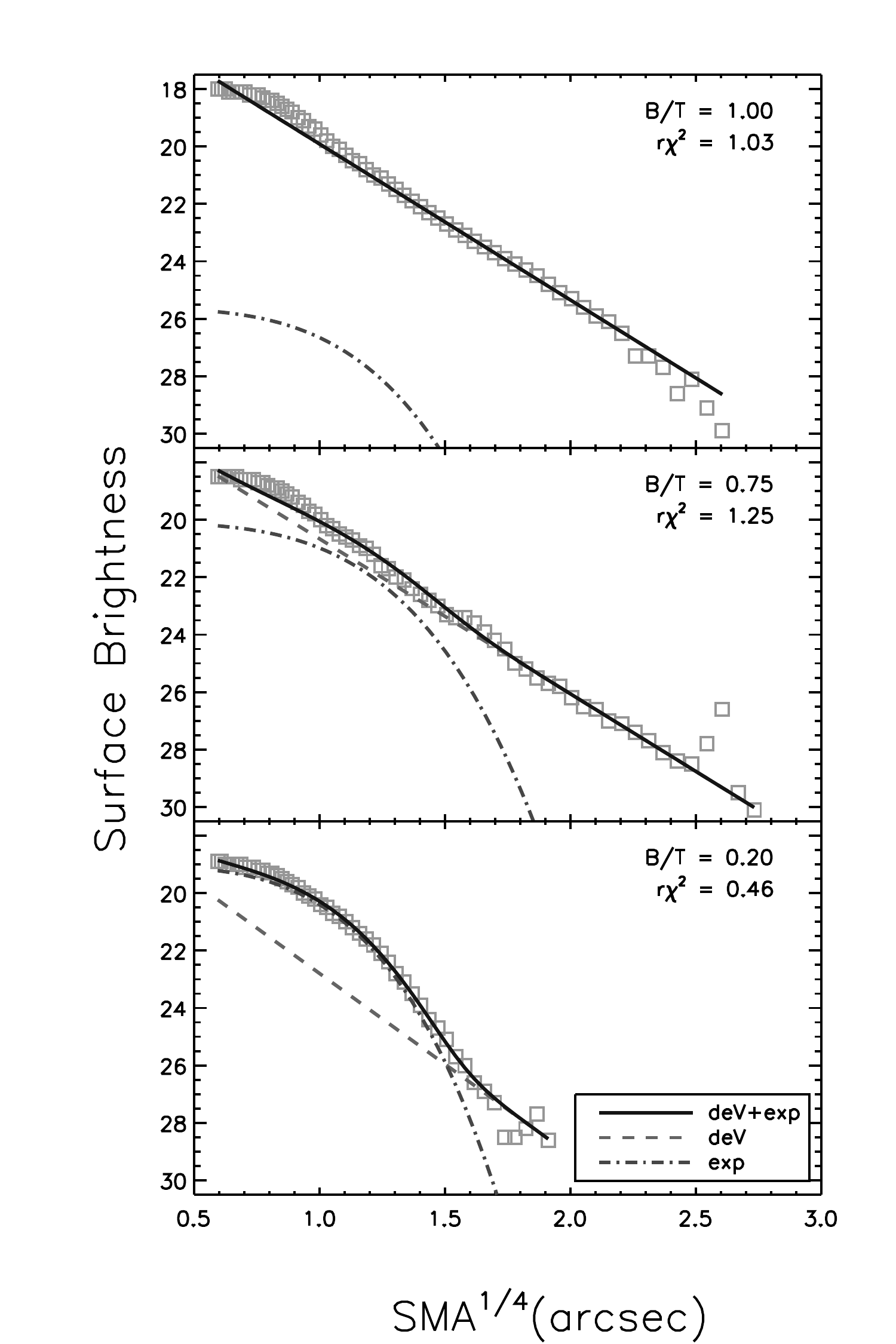}
\caption{Galaxy radial profiles and the fitting results are presented. The open rectangles indicate that the ellipsoid is fitted successfully for the annulus by the \texttt{ellipse} task. The best fit of the surface brightness profile is depicted using solid lines over 
the data points.  The de Vaucouleurs profile 
and the exponential profile comprising the best fit are also presented, in dashed lines and dashed-dotted lines, respectively. This figure 
shows that the galaxy profiles are robustly measured to $\mu_{r'} \sim$ 30 mag/arcsec. \label{btfit_sample}}
}
\end{figure}

\begin{figure}[p]
\center{
\includegraphics[scale=0.25]{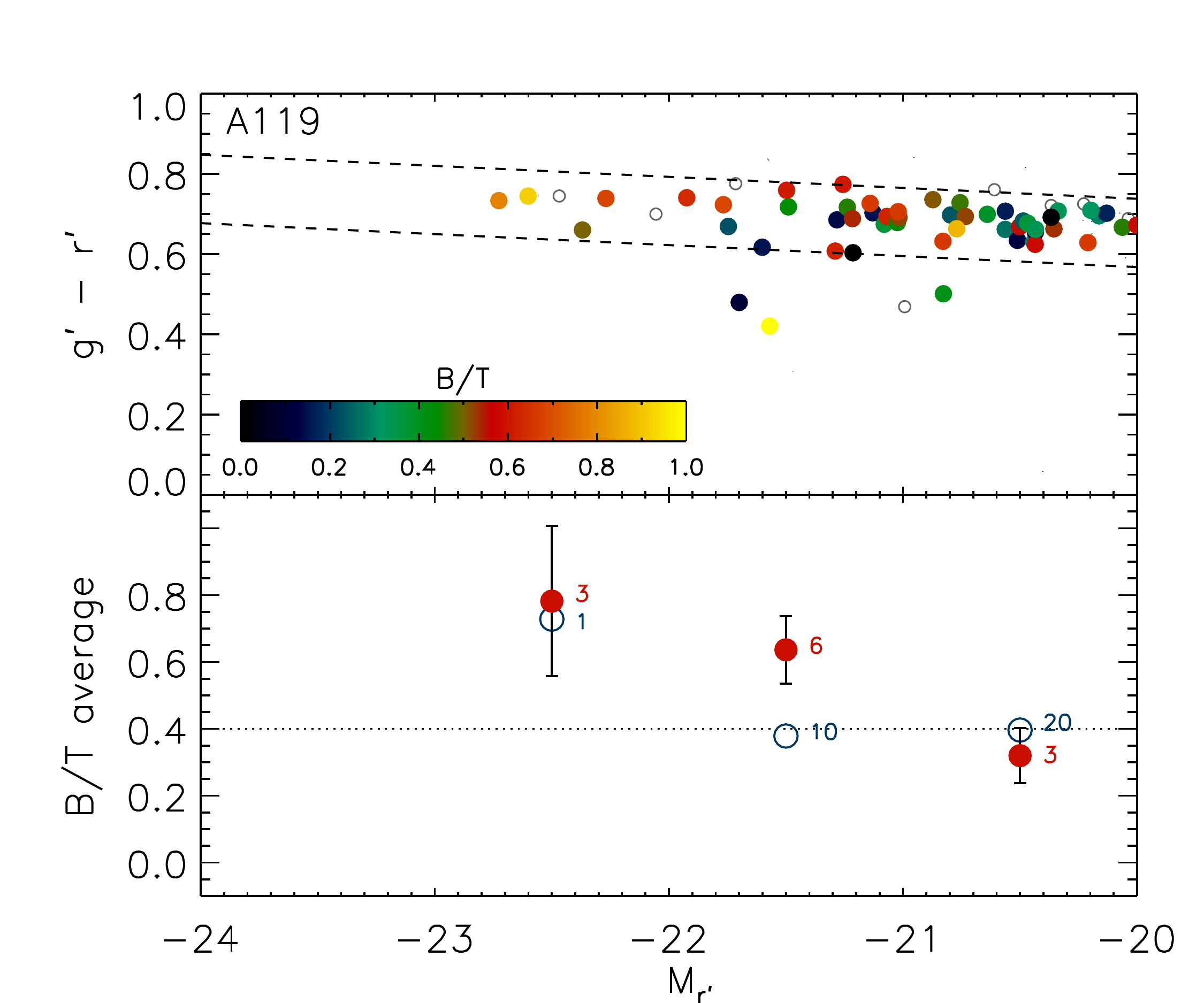}
\includegraphics[scale=0.25]{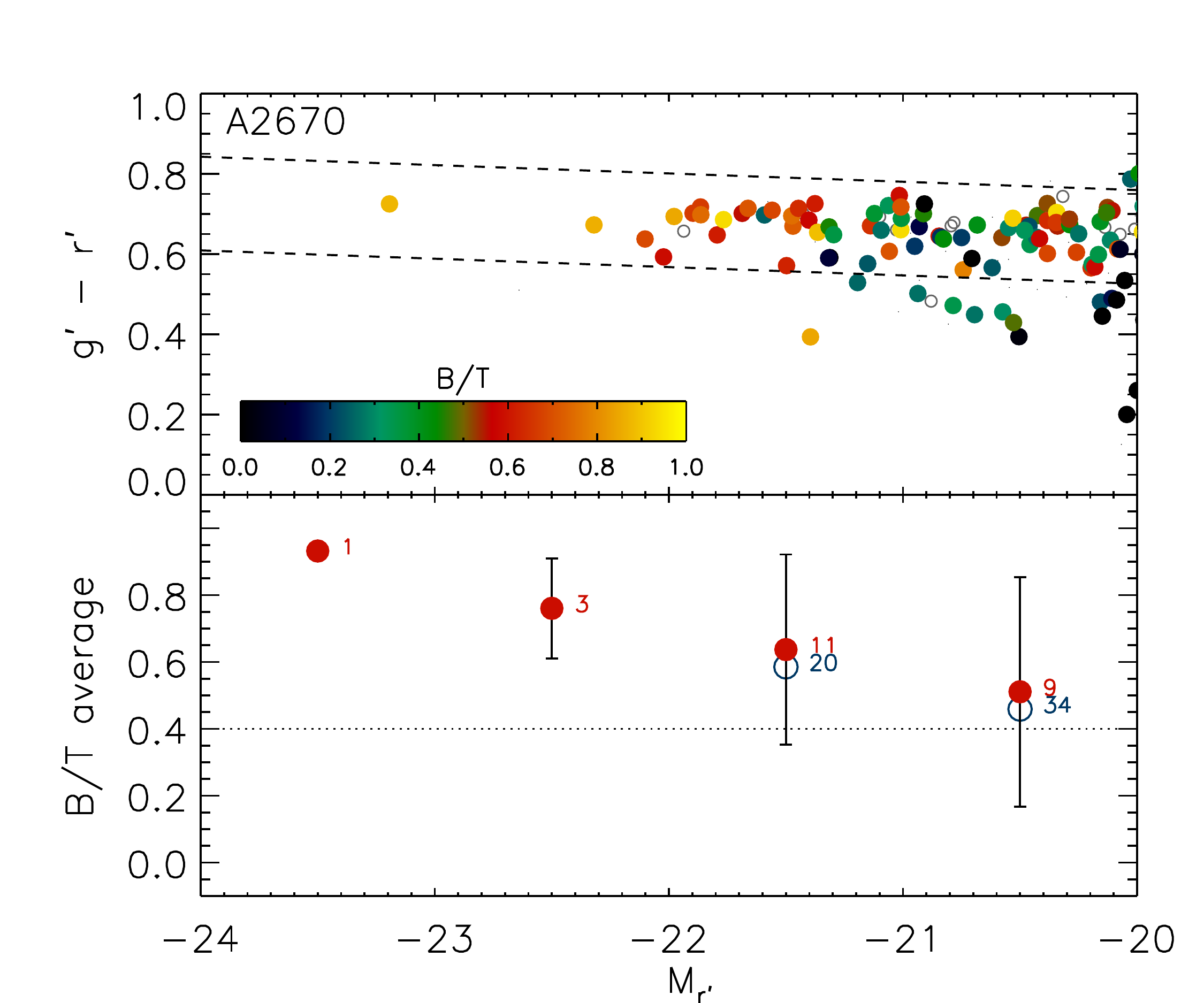}
\includegraphics[scale=0.25]{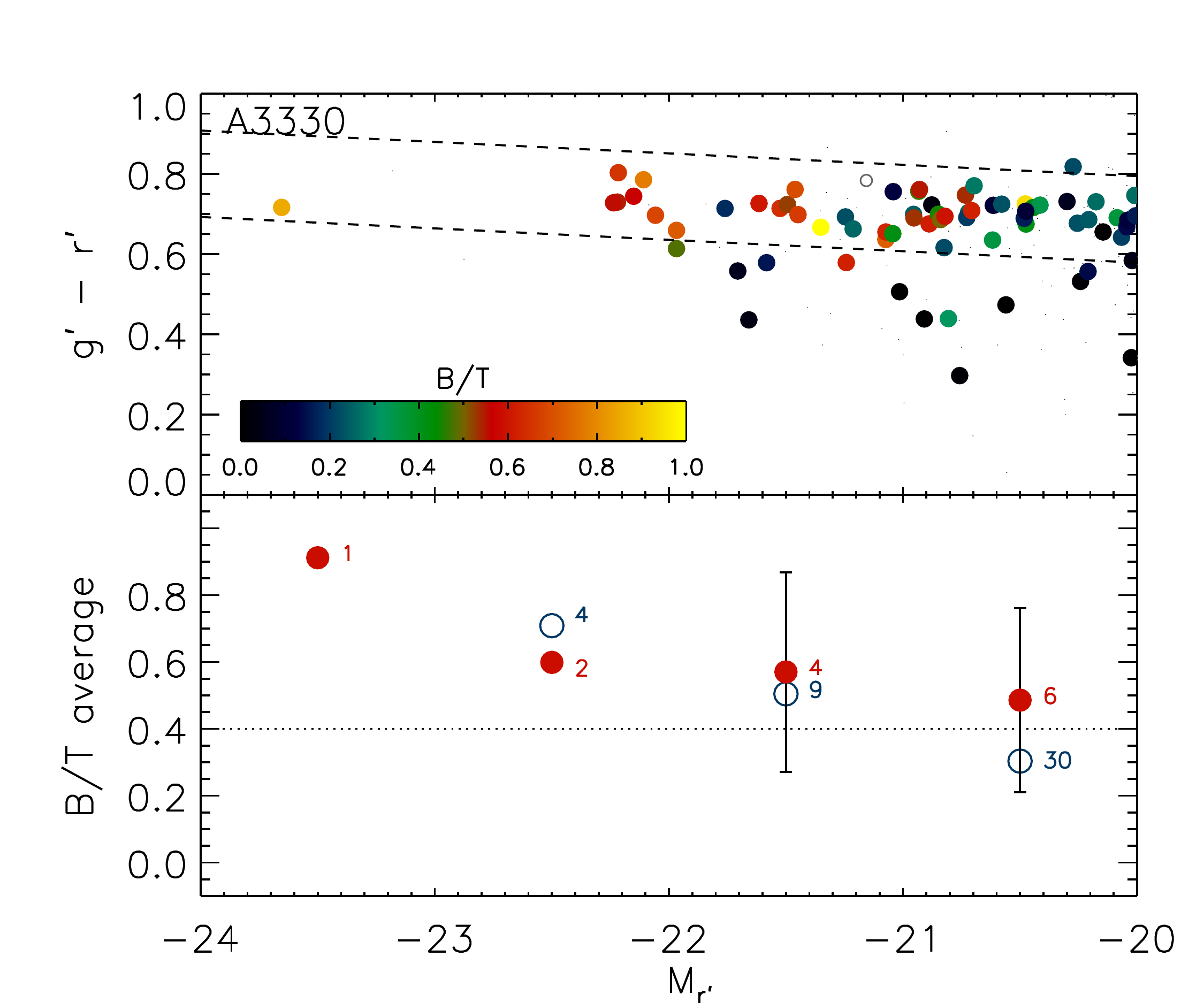}
\includegraphics[scale=0.25]{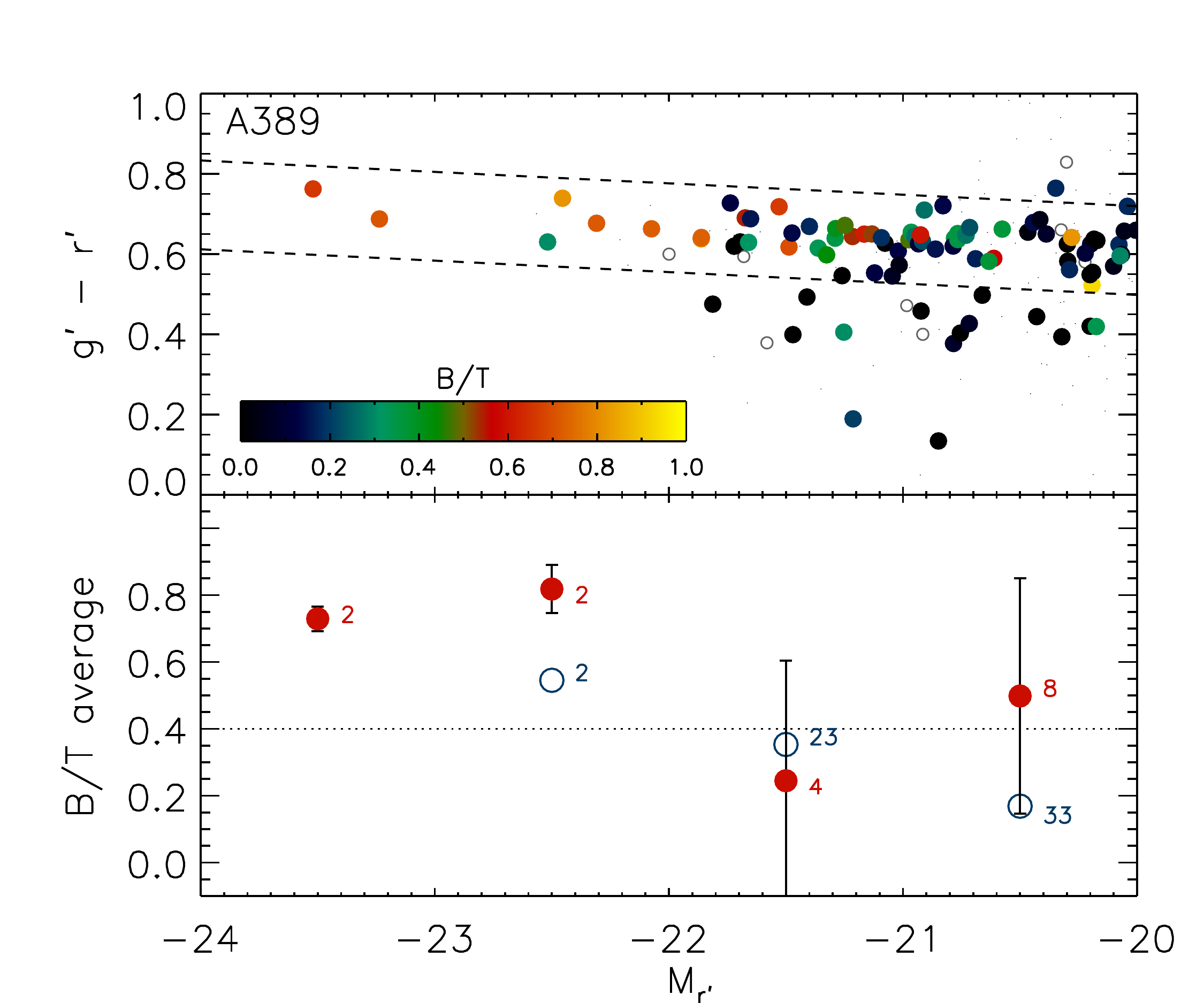}
\caption{The derived B/T values are presented for the spectroscopic member galaxies brighter than $M_{r'} = -20$ in each galaxy cluster. 
\textit{upper:} The B/T are expressed with color-coded filled circles in the optical CMR of each galaxy cluster. The dashed lines indicate color 
ranges of the red-sequence galaxies. The graphs confirm that bright red-sequence galaxies are mostly bulge-dominated systems with large B/T
values while the blue galaxies show small B/T values uniformly. \textit{bottom:} The averages B/T of post-merger galaxies among RS$_{sp}$ 
galaxies are plotted along the magnitude with the red-filled circles. The error bars show standard deviations of the mean values. Blue open circles 
are the average B/T of RS$_{sp}$ galaxies except the PM galaxies, presented for comparison. The number of galaxies counted for each average 
value is indicated beside the data point. We demarcate B/T $=$ 0.4 in the plots with the dotted lines, with which E/S0 galaxies are distinguished 
from spirals in this work. The graphs indicate that most of PM galaxies are bulge-dominated E/S0 galaxies.  \label{graph_comb_pm}}
}
\end{figure}

\begin{figure*}
\center{
\includegraphics[scale=0.7]{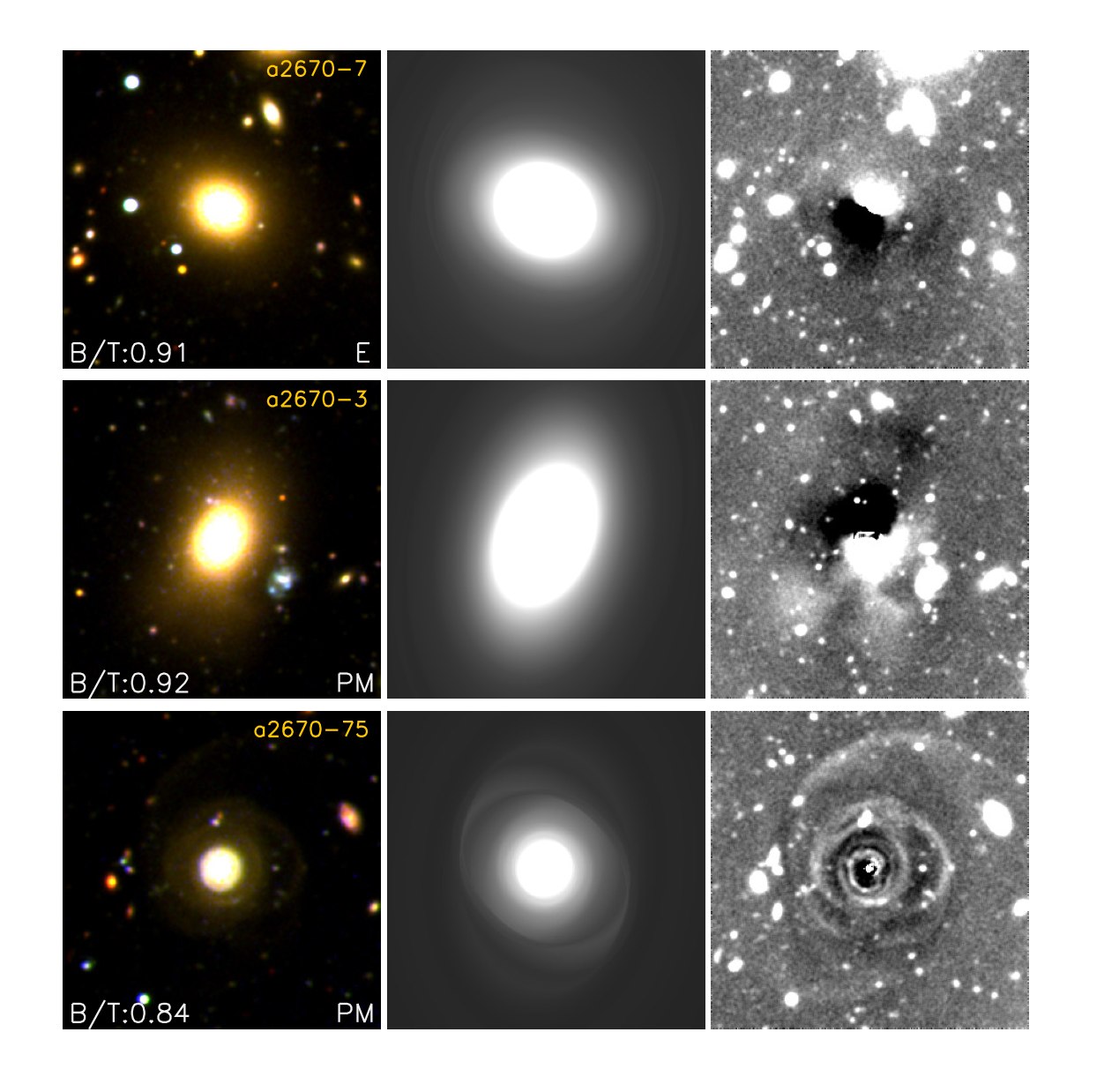}
\caption{Pseudo-color images, two-dimensional galaxy models and model-subtracted residual images of galaxies are presented from left to right 
in the same scale for each galaxy. The images of an early-type galaxy are presented in the top panels for comparison. The residual in the middle panel shows an 
example of discontinuous (in other words, structured) halo of an early-type galaxy while the bottom panel exhibits an obvious merger remnant 
with a long tidal tail and the spiral structures.  \label{resi_sample}}
}
\end{figure*}

\section{Morphological Examination}
\label{sec:mi}

\subsection{Bulge-to-total ratios and residual images}

Galaxy bulge-to-total ratios (B/T) were measured for the spectroscopic members of the galaxy clusters. Ratios are derived from the radial surface
brightness profiles of the galaxies, as measured by the {\tt ellipse} task in IRAF. Because the cores of bright galaxies are usually saturated in the 
deep images, we merged the two profiles from the shallow image and the deep image for the central part and the outer part of a galaxy, 
respectively. The conjunction is made at the radius where both profiles show the same surface brightness.

We found the best fits of the galaxy profiles to the composite galaxy model of the de Vaucouleurs profile (Sersic index, n = 4) and the exponential
profile. The galaxy surface brightness model can be expressed as, 
\begin{equation}
\label{sbeq}
I(R) = I_{e}~exp(c\times((R/R_{e})^{1/4} - 1))~+~I_{d}~exp(-R/R_{d}).
\end{equation}
Here, $R_{e}$ is the effective radius of the de Vaucouleurs profile and $I_{e}$ is the surface brightness at $R_{e}$. The constant, $c$, is set as a
free parameter to be determined. $R_{d}$ is the disk scale length and the $I_{d}$ is the surface brightness at $R_{d}$. 
The fitting is performed using the IDL routines in the MPFIT package \citep{mar08}. Usually such a fitting procedure is very sensitive to the initial 
values of the fitting parameters. To increase the success rate of the fitting, galaxy profiles are fitted with the de Vaucouleurs profile first. 
The preliminarily derived values of $I_{e}$, $R_{e}$ and $c$ are then used as an initial guess for the more complex main fitting with the 
equation~(\ref{sbeq}) and all the parameters are determined again as well as $I_{d}$ and $R_{d}$. The galaxy B/T is calculated using the following formula:

\begin{equation}
B/T = \frac{R_{e}^2 I_{e}}{R_{e}^2 I_{e}~+~0.28 R_{d}^2 I_{d}}
\end{equation}

We present examples of the galaxy profile fitting in Figure~\ref{btfit_sample}. The best fit of the surface brightness profile is drawn over the data 
points using the parameters derived from the fitting. The de Vaucouleurs profile and the exponential profile comprising the best fit are also 
presented. The figure shows that the galaxy profiles are robustly measured to $\mu_{r'} \sim$ 30 mag/arcsec.

Figure~\ref{graph_comb_pm} shows the result of B/T calculations for the cluster members. In the upper panel of each graph set, galaxies are 
expressed in different colors along their B/T values in the optical CMR. The high values of B/T of bright red-sequence galaxies and 
the low values of blue galaxies are consistent with the general expectation. In the bottom panels, average B/T values for RS$_{sp}$ are plotted 
along the absolute magnitude. As the upper panels have already suggested, it was found that the average B/T becomes smaller as the galaxies
become fainter.

We also computed the averages for the post-merger galaxies separately and presented them in red filled circles with the standard deviations. 
The results indicate that the average B/T values for post-merger galaxies are slightly higher than the average of the rest of the RS$_{sp}$ 
galaxies in the magnitude range.

We classified galaxies with B/T $>$ 0.4 as ellipticals and S0s following the criteria applied in \citet{som99}. The results show that a large fraction
of the featured galaxies are bulge-dominated systems as 77\% (10/13), 72\% (18/25), 77\% (10/13) and 59\% (10/17) for A119, A2670, A3330, 
and A389, respectively. This implies that possible contamination from disturbed spirals is not significant in our visual inspection.
As a result, the fractions of post-merger galaxies among the bulge-dominated systems (B/T $>$ 0.4) are about 40\% (10/25), 34\% (18/53), 36\% 
(10/28) and 48\% (10/21) for A119, A2670, A3330, and A389, respectively. On average, it is 38 $\pm$ 5\% in our cluster sample.
If we apply a more strict B/T criterion for bulge-dominated systems as B/T $>$ 0.6, the average of post-merger galaxy fractions increases to 
46 $\pm$ 8\%.

Two-dimensional galaxy surface brightness models are constructed with the result of \texttt{ellpse} fitting and are subtracted from the deep images
in order to check the features of the post-merger galaxies. The residual images of three RS$_{sp}$ galaxies with similar B/T values are shown in 
Figure~\ref{resi_sample}. At the same time, the images of three galaxies demonstrate the robustness of our B/T calculation for early-type galaxies.
Pseudo-color images, two-dimensional galaxy models and model-subtracted residual images are presented from left to right. An image of an 
early-type galaxy is presented in the top panels for comparison. Middle and bottom panels show examples of post-merger galaxies showing 
asymmetric, disturbed structures in their residual images.

\subsection{Asymmetry}
\begin{figure*}
\center{
\includegraphics[scale=0.6]{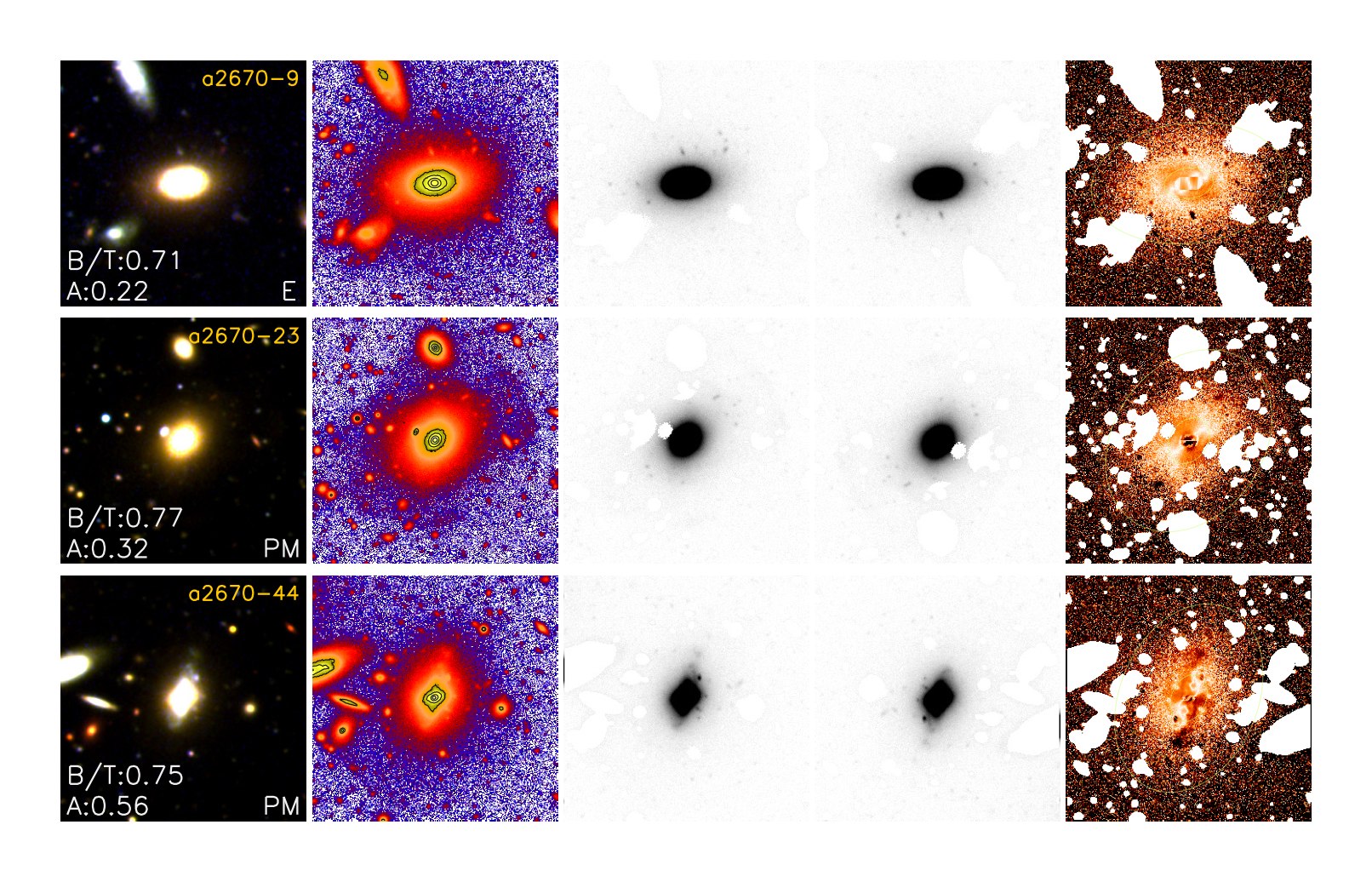}
\caption{Asymmetry. A color composite image (first), surface brightness map (second), r band deep image (third), the same deep image rotated 
180 degrees (fourth), and a normalized residual image (fifth). The green ellipses in the last column indicate 0.5$R_{Petro}$ and 2$R_{Petro}$ of 
each galaxy in which $\mathcal{A}$ was measured. \label{asymm_sample}}
}
\end{figure*}

\begin{figure}
\center{
\includegraphics[scale=0.5]{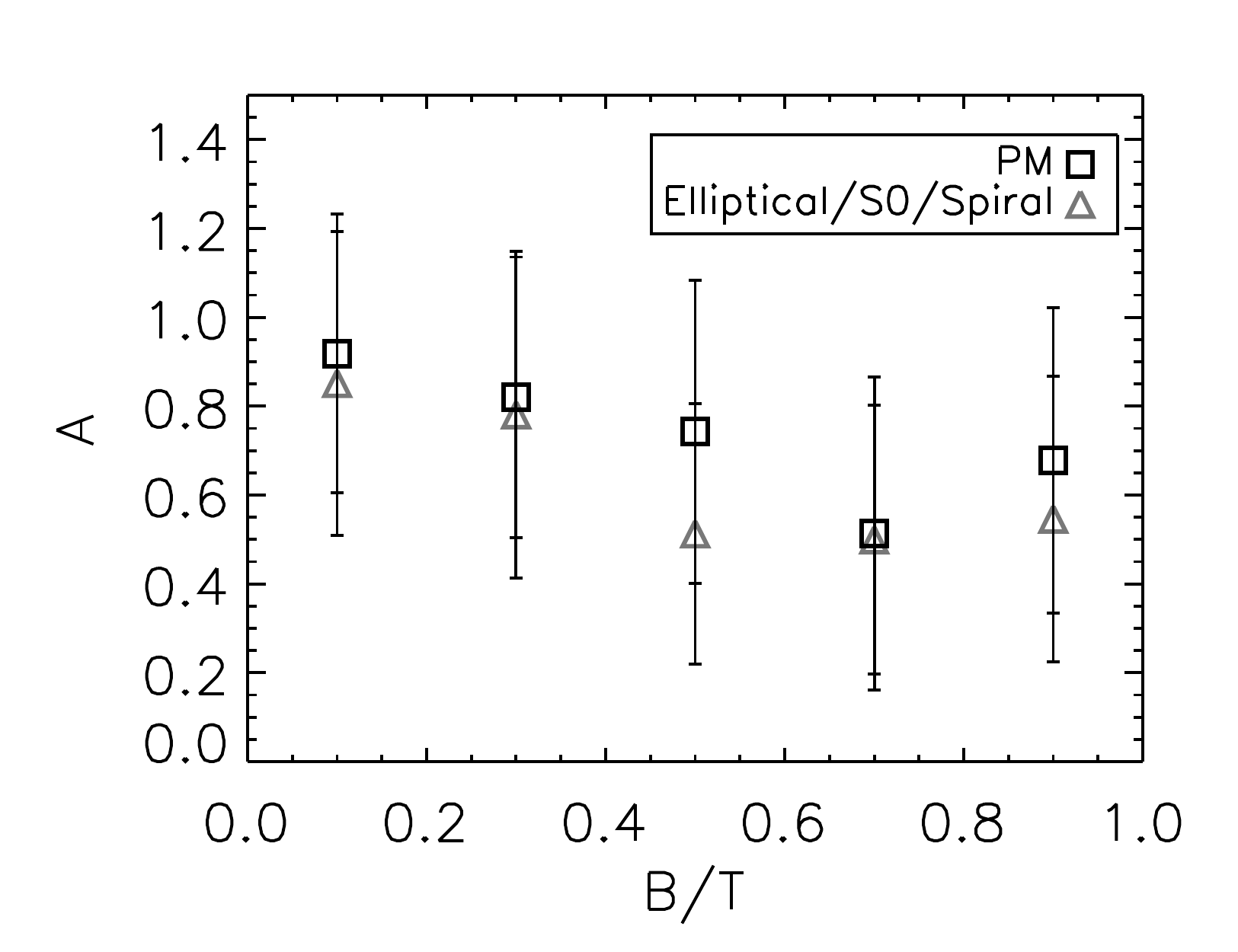}
\caption{Asymmetry distribution of red-sequence member galaxies along their B/T. Compared to normal elliptical and spiral galaxies, PM galaxies 
showed a slightly larger $\mathcal{A}$ on average. \label{asymm}}
}
\end{figure}

Using the $r'$ band deep images, we measured the asymmetry index, $\mathcal{A}$ for the cluster members. The measurements were made 
within the area between 0.5$R_{Petro}$ and 2$R_{Petro}$ of each galaxy avoiding galaxy centers usually saturated.
 
The $\mathcal{A}$ can be expressed as,

\begin{equation}
\mathcal{A} = \frac{\sum_{i,j}{| I(i,j) - I_{180}(i,j) |}}{\sum_{i,j}{|I(i,j)|}}
\end{equation}
where $I$ is an original image and the $I_{180}$ is the image rotated by 180\arcdeg.
We demonstrate the calculation process with some examples of RS$_{sp}$ galaxies showing similar B/T values in Figure~\ref{asymm_sample}.
The first column shows pseudo-color images and the second one displays $r'$ band surface brightness map of the galaxies.
The third and fourth images show the original deep image and the same image rotated 180 degrees. The rotated images are
subtracted from the original images and the normalize by the original images. The last column presents the normalized
residual images with green ellipses indicating 0.5$R_{Petro}$ and 2$R_{Petro}$ of each galaxy in which the $\mathcal{A}$ was
measured. Prior to the calculations, all nearby objects around the galaxies were masked from the images as shown in the images with white 
masks.

The average $\mathcal{A}$ of normal elliptical and spiral galaxies and the post-merger galaxies are plotted along their B/T values in 
Figure~\ref{asymm}. While the figure confirms that late-type galaxies show large asymmetry values compared to early-type galaxies, we found 
that the asymmetry in the post-mergers are slightly larger than that of normal early-type galaxies on average. Although the result corresponds to 
simple expectation, however, the averages of $\mathcal{A}$ contain a large spread of the values as expressed with error bars in the figure.
We suppose that the post-merger features in such a faint surface brightness level of our $r'$ band data were not easily recognizable 
with the asymmetry index, $\mathcal{A}$.

\section{Discussion}
\label{sec:disc}

We investigated the post-merger signatures of red-sequence galaxies in rich Abell clusters at $z \lesssim$ 0.1. Deep images in $u'$, $g'$, $r'$ 
and medium-resolution galaxy spectra were taken for A119, A2670, A3330 and A389 with a MOSAIC 2 CCD and a Hydra MOS mounted on the 
Blanco 4-m telescope at CTIO. Post-merger signatures were identified by visual inspection of their disturbed features, e.g., asymmetric structures, 
faint features, discontinuous halo structures, rings and dust lanes. Most ($\sim$ 71\%) of the featured galaxies were found to be bulge-dominated
systems (E/S0 galaxies) from the radial surface brightness profile fitting. On average, the asymmetry of the post-merger galaxies is always slightly
larger than the mean of normal galaxies within the same bulge-to-total ratio range.

We noted that cluster red-sequence galaxies went through galaxy mergers until the recent epoch. In this work, $\sim$ 25\% of the bright 
($M_r <  -20$) cluster red-sequence (RS$_{sp}$) galaxies exhibited post-merger signatures in four target clusters consistently. 
The fraction increases to $\sim$ 38\% with the bulge-dominated RS$_{sp}$ galaxies (B/T $>$ 0.4). In addition, brighter galaxy samples show 
higher fractions of post-merger signatures in the clusters.

\begin{deluxetable}{cccc}
\tablecolumns{4}
\tabletypesize{\footnotesize}
\tablewidth{0pt}
\tablecaption{Comparisons between cluster and field \label{compfield}}
\tablehead{
\colhead{} & \colhead{Class} & \colhead{Cluster} & \colhead{Field\tablenotemark{a}}
}
\startdata
 & PM & 25 $\pm$ 3\% & 35\% \\
Red\tablenotemark{b} & I & 5 $\pm$ 1\% & 18\% \\
 & Total & 30 $\pm$ 4\% & 53\% \\[3pt]
\hline \\[-3pt]
 & PM & 38 $\pm$ 5\% & 49\% \\
Bulge-dominated\tablenotemark{c}& I & 4 $\pm$ 1\% & 21\% \\
 & Total & 42 $\pm$ 6\% & 70\%
\enddata

\tablenotetext{a}{The fractions for the field environment were adopted from \citet{van05}.}
\tablenotetext{b}{Fractions for the cluster are derived with the RS$_{sp}$ galaxies in this paper, while the field red galaxies are denoted with 
$B - R$ colors.} 
\tablenotetext{c}{Fractions with only bulge-dominated galaxies among the red galaxies. For the cluster, galaxies with B/T $>$ 0.4 are included 
while visually classified E/S0 galaxies are considered for the field.} 
\end{deluxetable}

In Table~\ref{compfield}, we compared the average of fractions from the four Abell clusters with the result of field environment from \citet{van05} . 
The clusters and field environments were compared using two subsets of galaxy samples, red galaxies and bulge-dominated red galaxies. 
Our RS$_{sp}$ galaxies were compared with the luminous red galaxies selected using $B-R$ colors in \citet{van05}, and the bulge-dominated 
galaxies (B/T $>$ 0.4) were compared with the visually classified field E/S0 galaxies in the same paper. 
We presented the fractions of post-merger galaxies and interacting galaxies separately as well as the combined fractions which indicate 
the fraction of disturbed galaxies related to past or ongoing galaxy mergers. The van Dokkum sample does come with spectroscopic redshift; 
spectroscopic redshifts were available only on 5 galaxies out of 126 and so no k-correction was applied. Based on a small subsample he argues 
that his galaxies are roughly at redshift of 0.1, which is comparable to ours. The colors of his bulge-dominated galaxies are also comparable to 
ours after proper color conversions using \citet{lup05}.

The table demonstrates that the overall fractions in clusters are lower than those in the field. We find it difficult to understand the ``interacting" 
galaxies in these clusters. They appear to be beginning their interaction/merger, and so we classified them as ``interacting" or pre-merger. 
Considering that interacting systems with small companions are likely missed in our search, it is likely that our fraction for interacting galaxies is 
a hard lower limit. For example, some of our ``faint companion galaxies" (which were excluded in our analysis) could be ``interacting". Direct 
satellite-satellite mergers are supposed to be extremely rare in large halo environments, and so it is difficult to understand them.

The fractions of post-merger galaxies are not very different between the cluster and the field. This is not compatible with a simple theoretical 
expectation based on merger timescale as a function of relative speeds of galaxies. Therefore, we suggest that many massive red-sequence 
galaxies formed in a less dense environment through galaxy mergers and entered into the cluster through dark matter halo mergers.

An important caveat exists on this comparison between van Dokkum's work and ours. Merger timescales and thus probability heavily depends 
on mass ratio of colliding systems, but it is very difficult to extract the mass ratio information from observed images. We thought it would be the 
best attempt to use similar instruments for similar depth (exposure) on galaxies at similar distances. But even in this case, we are not free from a 
possible environmental (cluster vs field) dependence on mass ratios of galaxy mergers.

\begin{figure}
\center{
\includegraphics[scale=0.55]{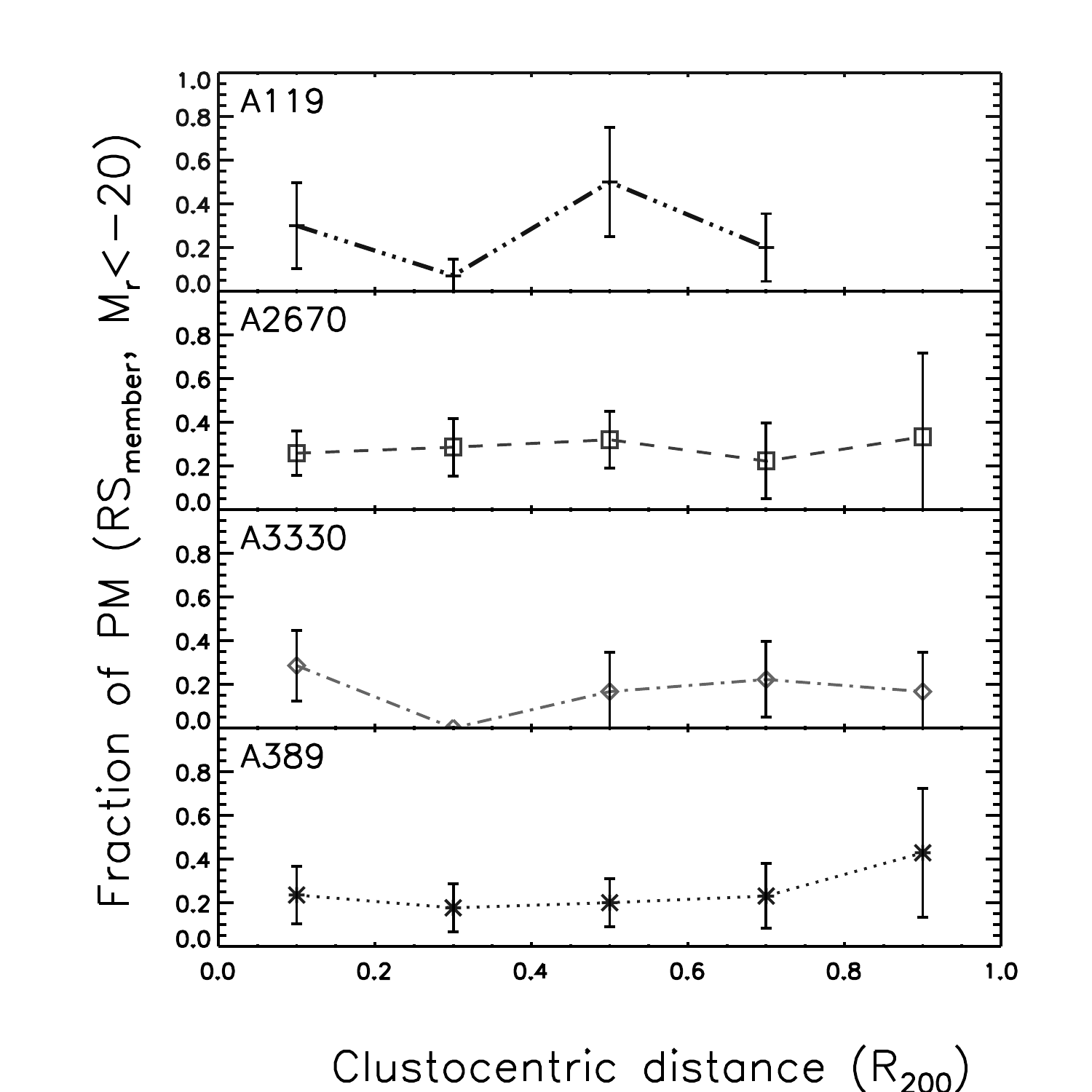}
\caption{Fractions of PM galaxies along the projected distances from the BCGs. The size of distance bin is 0.2$R_{200}$ in each galaxy cluster. 
Contrary to the general assumption, which would predict more frequent mergers at the outskirts of a cluster, the fraction of PM galaxies does not 
change much along the clustocentric distances. \label{distance}}
}
\end{figure}

The radial distribution of the post-merger galaxies in the clusters provides another hint supporting this scenario. The fractions of post-merger 
galaxies are derived along the projected distance from the BCG in the scale of $R_{200}$ of each galaxy cluster, as shown in 
Figure~\ref{distance}. We found that the fractions do not change much along the clustocentric distances. In situ mergers can scarcely explain 
this result because it is difficult to expect frequent mergers in the central region of clusters compared to the outskirts. 
This implies again that a large fraction of massive red-sequence galaxies with post-merger signatures likely merged in a less dense environment 
and then fell into the central region of the cluster.

Is it possible for a galaxy to maintain the post-merger features produced in the outskirts until the galaxy arrives at the cluster center? It has been 
suggested for a long time that tidally ejected materials during galaxy mergers may return over many Gyr \citep{her92, hib95}. \citet{jen08} also 
proposed that the features of interacting galaxies can be observable for up to $\sim$ 2 Gyr by applying stellar population models to the simulated 
equal-mass gas-rich disc mergers. We note that the duration of merger feature is highly sensitive to galaxy type, merger geometry,
mass ratio, and last but not least the observing surface brightness limit\footnote[1]{~The suggestion of 2 Gyr visibility of merger 
features by \citet{jen08} was based on the surface brightness, $\mu =$ 25 assumption. The visibility time is likely longer in our case because our
imaging was deeper.}.

The dynamical friction timescale of a satellite galaxy to the central cluster halo would be another parameter to compare with the observable 
timescale of post-merger features. In this work the dynamical friction timescale is defined as the timescale between the epochs of when the halo 
of satellite galaxy is started to merge into the cluster halo at the cluster virial radius and when the satellite galaxy have finally merged into the 
BCG in the model. If a dynamical friction timescale for a galaxy is much longer than the timescale of the post-merger features, our results should
be interpreted in a different light. By taking advantage of a semi-analytic approach, we obtained the relationship between the merger mass ratio 
and the merger timescale of galaxies to the BCG. Our semi-analytic model (SAM) is based on N-body backbone dark matter merger trees and 
physical ingredients that govern the baryonic properties of galaxies. The merger timescales in the model are calculated using the formulae of
\citet{jia08}, which provides a modified form of Chandrasekhar's formula. Given that Chandrasekhar's formula generally underestimates merger 
timescales, \citet{jia08} fits the formula into merger timescales derived from hydrodynamic simulations \citep[see also][]{kim11}.

\begin{figure}
\center{
\includegraphics[scale=0.45]{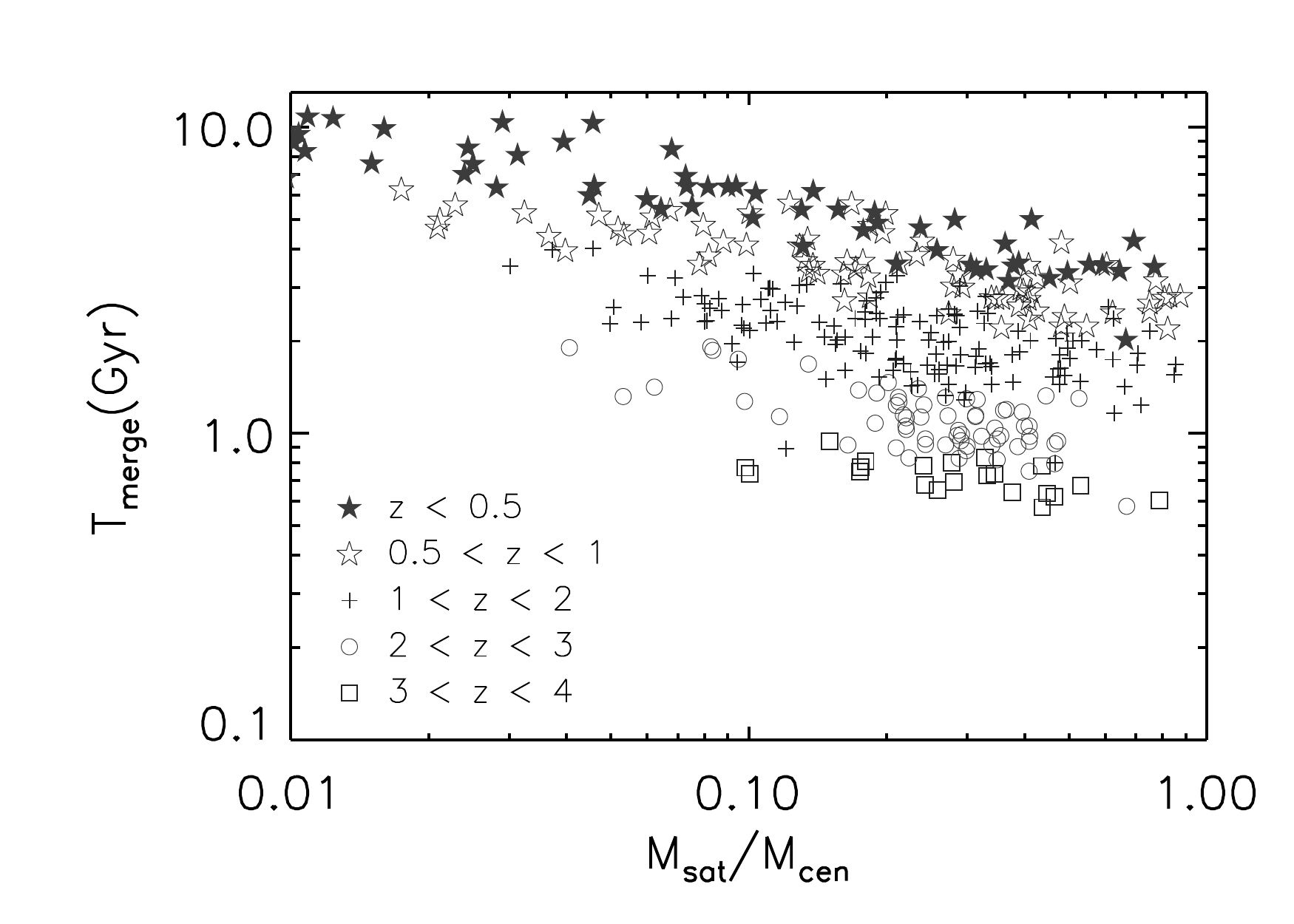}
\caption{Merging timescale of galaxies in a cluster of 5$\times10^{14}$M$_{\sun}$ of our SAM along the mass ratio. If we assume that a central 
galaxy is the BCG, bright satellite galaxies (M$_{sat} >$ 0.25M$_{cen}$) merge into the BCG within $\sim$ 4 Gyr at $z <$ 0.5.  
\label{tmerge}}
}
\end{figure}

We selected a cluster in the model which has a cluster mass, 5$\times10^{14}$M$_{\sun}$, comparable to our cluster samples. 
Figure~\ref{tmerge} shows the merging timescales of all satellite galaxy mergers along the mass ratio between the BCG and the satellite, 
consequently forming a cluster of 5$\times10^{14}$M$_{\sun}$ at present. The redshift of each merger was expressed using different symbols, 
showing that the merger timescale was shorter at a higher redshift for the same mass ratio in the model. This arises because the size of 
the cluster was smaller at an early epoch. We can assume that the oldest observable post-merger features in our target clusters are produced 
at $z \sim$ 0.5, where the difference in the lookback time from $z \sim$ 0.1 is about 4 Gyr. The figure shows that the dynamical friction timescale 
of a satellite galaxy as massive as M$_{sat}$/M$_{cen} >$ 0.25 will be $\lesssim$ 4 Gyr below the redshift, $z =$ 0.5. If we assume that the 
magnitude difference between the BCG and the second BCG is 1 magnitude, then the mass ratio would be $\sim$ 0.4 (M$_{sat}$:M$_{cen} =$
1:2.5). The figure indicates that then the second BCG will merge into the BCG within 3 Gyr at $z <$ 0.5. Therefore, it appears that massive 
post-merger galaxies can fall into the cluster center and maintain their disturbed features, even in low redshift ($z <$ 0.5). 
The consistent fractions of the featured galaxies along the clustocentric distance most likely arise from on-going halo mergers which started 
during various epochs.

The average count of post-merger galaxies in our cluster samples was 17 $\pm$ 6 galaxies (13, 25, 13 and 17 galaxies for A119, A2670, A3330 
and A389, respectively). We also counted the number of halo mergers related to the current cluster halo in the SAM, to check whether a cluster 
can experience sufficient halo mergers to absorb massive galaxies since $z=$ 0.5. In the model of a 5$\times10^{14} $M$_{\sun}$ cluster, we 
found 21 satellite halos merging into the cluster at $z <$ 0.5, with galaxies brighter than $M_{r'} = -20$. It theoretically supports that a massive
cluster may have enough halo mergers to take recently merged galaxies in at recent epoch. In this preliminary analysis, we assumed that merger
features from the previous halo environment last roughly the same time as the subhalo's dynamical friction timescale. This is certainly an 
over-simplification. A more realistic calculation would require accurate merging halo mass ratios, merging galaxy mass ratios, and merger 
feature timescales throughout cluster evolution history.

In conclusion, we found that $\sim$ 42\% of massive, bulge-dominated red-sequence galaxies in galaxy clusters have continued their mass 
assembly through galaxy mergers until the recent epoch, as $\sim$ 70\% of the field galaxies have done. Although the fractions of post-merger
galaxies in clusters are lower than that in the field, it is still too high compared to the expectation from apparent fractions of galaxy interaction in
galaxy clusters. Therefore, we propose that most of those post-merger galaxies were assembled in a low-density region and fell into the current 
cluster via halo mergers. We have supported this scenario with theoretical predictions using a semi-analytical model.

We do not speculate the progenitors of the post-merger galaxies in this paper, regardless of whether they are remnants of dry mergers or wet
mergers. This will be investigated in depth in an upcoming paper which will discuss on the ultraviolet properties of the galaxies most likely related
to the residual star formation induced by galaxy mergers.

\

\acknowledgments

We thank Yujin Yang for helping us obtain the short exposure images of Abell 389.
Y. -K. Sheen is grateful to Knut Olsen, Francisco Valdes and Mike Fitzpatrick for their helpful comments and discussions on the MOSAIC 2 
and Hydra data. 
We acknowledge the support from the National Research Foundation of Korea to the Center for Galaxy Evolution Research and Doyak grant 
(No. 20090078756) and from the Korea Astronomy and Space Science Institute.



{\it Facilities:} \facility{MOSAIC 2, Hydra (CTIO)}


\begin{thebibliography}{}
\bibitem[Bell et al.(2004)]{bel04} Bell, E. F., Wolf, C., Meisenheimer, K., et al. 2004, \apj, 608, 752
\bibitem[Bertin \& Arnouts(1996)]{ber96} Bertin, E., \& Arnouts, S. 1996, \aaps, 117, 393
\bibitem[Bower, Lucey \& Ellis(1992)]{bow92} Bower, R. G., Lucey, J. R., \& Ellis, R. S. 1992, \mnras, 254, 601 
\bibitem[Bundy et al.(2009)]{bun09} Bundy, K., Fukugita, M., Ellis, R. S., et al. 2009, \apj, 697, 1369  
\bibitem[Carlberg et al.(1997)]{car97} Carlberg, R. G., Yee, H. K. C., \& Ellingson, E. 1997, \apj, 478, 462
\bibitem[de Ravel et al.(2009)]{der09} de Ravel, L., Le F$\grave{e}$vre, O., Tresse, L., et al. 2009, \aap, 498, 379
\bibitem[Dressler(1980)]{dre80} Dressler, A. 1980, \apj, 236, 351
\bibitem[Drory \& Alvarez(2008)]{dro08} Drory, N., \& Alvarez, M. 2008, \apj, 680, 41
\bibitem[Faber et al.(2007)]{fab07} Faber, S. M., Willmer, C. N. A., Wolf, C., et al. 2007, \apj, 665, 265
\bibitem[Gunn \& Gott(1972)]{gun72} Gunn, J. E., \& Gott, J. R. 1972, \apj, 176, 1
\bibitem[Hernquist \& Spergel(1992)]{her92} Hernquist, L., \& Spergel, D. N. 1992, \apj, 399, L117
\bibitem[Hibbard \& Mihos(1995)]{hib95} Hibbard, J. E., \& Mihos, J. C. 1995, \aj, 110, 140
\bibitem[Jannuzi, Claver \& Valdes(2003)]{jan03} Jannuzi, B. T.,
  Claver, J., \& Valdes, F. 2003, ``The NOAO Deep
  Wide-Field Survey MOSAIC Data Reduction'' 
(\url{http://www.noao.edu/noao/noaodeep/ReductionOpt/frames.html})
\bibitem[Jiang et al.(2008)]{jia08} Jiang, C. Y., Jing, Y. P., Faltenbacher, A., Lin, W. P., \& Li, C. 2008, \apj, 675, 1095
\bibitem[Kang \& Im(2009)]{kan09} Kang, E., \& Im, M. 2009, \apj, 691, L33
\bibitem[Kaviraj et al.(2007)]{kav07} Kaviraj, S., Schawinski, K., Devriendt, J. E. G., et al. 2007, \apjs, 173, 619
\bibitem[Kaviraj(2010a)]{kav10a} Kaviraj, S., 2010a, \mnras, 406, 382
\bibitem[Kaviraj(2010b)]{kav10b} Kaviraj, S., 2010b, \mnras, 408, 170
\bibitem[Kimm, Yi, \& Khochfar(2011)]{kim11} Kimm, T., Yi, S. K., \& Khochfar, S. 2011, \apj, 729, 11
\bibitem[Kodama et al.(2007)]{kod07} Kodama, T., Tanaka, I., Kajisawa, M., et al. 2007, \mnras, 377, 1717
\bibitem[Lotz et al.(2008)]{jen08} Lotz, J. M., Jonsson, P., Cox, T. J., \& Primack, J. R. 2008, \mnras, 391, 1137
\bibitem[Lupton(2005)]{lup05} Lupton, R. 2005, ``Transformation between SDSS magnitudes and $UBVR_{c}I_{c}$" 
(\url{http://www.sdss.org/dr7/algorithms/sdssUBVRITransform.html\\\#Lupton2005})
\bibitem[Markwardt(2008)]{mar08} Markwardt, C. B. 2008, ASP conference series, Vol. 411, p. 251
\bibitem[Schlegel et al.(1998)]{sch98} Schlege, D. J., Finkbeiner, D. P., \& Davis, M. 1998, \apj, 500, 525
\bibitem[Smith et al.(2003)]{smi03} Smith, J. A., Tucker, D. L., Allam, S. S., \& Rodgers, C. T. 2003, \aj, 126, 2037
\bibitem[Somerville \& Primack(1999)]{som99} Somerville, R. S., \& Primack, J. R. 1999, \mnras, 310, 1087
\bibitem[Stanford, Eisenhardt \& Dickinson(1998)]{sta98} Stanford, S. A., Eisenhardt, P. R., \& Dickinson, M. 1998, \apj, 492, 461
\bibitem[Tanaka et al.(2005)]{tan05} Tanaka, M., Kodama, T., Arimoto, N., et al. 2005, \mnras, 362, 268 
\bibitem[Tonry \& Davis(1979)]{ton79} Tonry, J., \& Davis, M. 1979, \aj, 84, 1511
\bibitem[Valdes(2000)]{val00} Valdes, F. 2000, ``Creating a Mosaic World Coordinate System'' 
(\url{http://iraf.noao.edu/projects/ccdmosaic/astrometry/astrom.html})
\bibitem[van Dokkum(2001)]{van01} van Dokkum, P. G. 2001, \pasp, 113, 1420
\bibitem[van Dokkum(2005)]{van05} van Dokkum, P. G. 2005, \aj, 130, 2647
\bibitem[Yi et al.(2011)]{yi11} Yi, S. K., Lee, J., Sheen, Y. -K., et al. 2011, \apjs, 195, 22
\end{thebibliography}
\end{document}